\documentclass[11pt]{article}
\usepackage{amsmath} 
\usepackage[dvips]{graphics}
\usepackage{amscd}     \usepackage{amsxtra}
\usepackage{upref}     \usepackage{amsthm}
\usepackage{amssymb}   
\usepackage{amsfonts}
\begin{document}

\title{{\bf
 A nonlinear prerelativistic approach to mathematical representation
of vacuum electromagnetism.}}

\author{{\bf Stoil Donev}\footnote{e-mail:sdonev@inrne.bas.bg},
{\bf Maria Tashkova}, \\
Institute for Nuclear
Research and Nuclear Energy,\\ Bulg.Acad.Sci., 1784 Sofia,
blvd.Tzarigradsko chaussee 72, Bulgaria\\}
\date{}
\maketitle

\begin{abstract}
This paper presents an alternative to the Maxwell vacuum equations
pre-relativistic approach to description of electromagnetic field objects. Our
view is based on the understanding that the corresponding differential
equations should be dynamical in nature and the physical relations represented
by them should represent local stress-energy-momentum balance relations. Such a
view does not go along with the classical assumption for local recognizability
of the electric and magnetic constituents $\mathbf{E}$ and $\mathbf{B}$ as
time-stable and space propagationg subsystems of the field objects. The
corresponding reconsideration brought us to the assumption, that the two
couples $\mathbf{(E,B)}$ and $\mathbf{(-B,E)}$ are much more adequate in this
respect: free electromagnetic field objects exist in a permanent propagation
with the fundamental velocity {\it c}, so each of its recognizable subsystems
should be able to carry momentum, and neither $\mathbf{E}$ nor $\mathbf{B}$ are
able to do this separately, while each of the couples $\mathbf{(E,B;-B,E)}$ is
able to do this, but only in presence of the other. Therefore, the necessary
internal local dynamics, admissible changes, time stability and recognizability
during space propagation should be viewed in terms of $\mathbf{(E,B)}$ and
$\mathbf{(-B,E)}$ and their mutually compatible changes.
 \end{abstract}


\begin{flushright}
{\it One must consider the source of nonlinearity in a field theory \\by
starting with {\bf physically reasonable assumptions}, ..., rather\\ than merely
generalizing the purely mathematical form of the\\ lagrangian or field
equations} [10, page 6].

D.H.Delphenich
\end{flushright}
\section{Introduction: General Notions about Physical Objects and
\\Interactions}
In correspondence with  modern theoretical view on classical fields we
consider time dependent and space propagating electromagnetic fields as flows
of spatially finite physical entities which have been called in the early 20th
century {\it photons}. Since the efforts made during the past years to find
appropriate in this respect nonlinearizations of Maxwell vacuum equations
[1],[2],[3],[4],[5],[6],[7],[8],[9],[10],\newline[11],[12],[13],[14],[15],
and the seriously developed quantum theory have not resulted so far in
appropriate, from our viewpoint, description of time stable entities of
electromagnetic field nature with finite spatial support, in particular,
adequate theoretical image of single photons as spatially finite physical
objects with internal dynamical structure, explaining in such a way their time
stability and spin nature, we decided to look back to the rudiments of the
electromagnetic theory trying to reconsider its assumptions in order to come to
equations giving appropriate solutions, in particular, solutions with spatially
finite carrier at every moment of their existence and space-propagating as a
whole, keeping, of course, their physical identity. This reconsideration
brought us to the necessity to admit that introducing as clear as possible
notions about the concepts {\it physical object} and {\it physical interaction}
is strongly needed. Here we briefly present the view on these concepts that we
shall follow throughout the paper.

When we speak about physical objects, e.g. classical
particles, solid bodies, elementary particles, fields, etc., we always suppose
that some {\it definite properties of the object under consideration do not
change during its time-evolution under the influence of the existing
environment}.  The availability of such time-stable features of any physical
object makes it {\it recognizable} among the other physical objects, on one
hand, and  guarantees its proper {\it identification} during its existence in
time, on the other hand. Without such an availability of constant in time
properties (features), which are due to the object's resistance and surviving
abilities, we could hardly speak about objects and knowledge at all. So, for
example, two classical mass particles together with the associated to them
gravitational fields survive under the mutual influence of their gravitational
fields through changing their states of motion: change of state compensates the
consequences of the violated dynamical equilibrium that each of the two
particles had been established with the physical environment before the two
gravitational fields have begun perturbating each other.

The above view implies that three kinds of quantities will be necessary to
describe as fully as possible the existence and the evolution of a given
physical object:

	1. {\bf Proper (identifying) characteristics}, i.e. quantities which
do NOT change during the entire existence of the object. The availability of
such quantities allows to distinguish a physical object among the other ones.

	2. {\bf Kinematical characteristics}, i.e. quantities, which describe
the allowed space-time evolution, where "allowed" means {\it consistent with
the constancy of the identifying characteristics}.

	3. {\bf Dynamical characteristics}, i.e. quantities, describing
surviving abilities through appropriate gaining and losing during existence.

Some of the dynamical characteristics must have the following two important
properties:  they are in a definite degree {\bf universal}, i.e., a class of
physical objects (may be all physical objects) carry nonzero value of them (e.g.
energy-momentum), and they are {\bf conservative}, i.e., they may just be
transferred from one physical object to another (in various forms) but no loss
is allowed.

Hence, the evolution of a physical object subject to bearable/acceptable
exterior influence (perturbation), coming from the existing environment,
has three aspects:

	1. {\it constancy of the proper (identifying) characteristics},

	2. {\it allowed kinematical evolution},

	3. {\it appropriate exchange of dynamical quantities with the physical
environment guaranteing object's time stability}.

Moreover, if the physical object under study is space-extended
(continuous) and demonstrates internal structure and dynamics, i.e., available
interaction of time-stable subsystems, it should be described by a number of
many-component mathematical object, e.g., vector valued differential form,
clearly, we must consider this {\it internal exchange} of some dynamical
characteristics among the various subsystems of the object as essential feature
determining in a definite extent object's integral appearence.

The above features suggest that the dynamical equations, describing locally
the evolution of the object, may come from giving an explicit form of the
quantities controlling the local internal and external exchange processes,
i.e. writing down corresponding local balance equations.  Hence, denoting
the local quantities that describe the external exchange processes by $Q_i,
i=1,2,\dots$, the object should be considered to be $Q_i$-free,
$i=1,2,\dots$, if the corresponding integral values are time constant,  which
can be achieved only if $Q_i$ obey differential equations presenting
appropriately (implicitly or explicitly) corresponding local versions of the
conservation laws (continuity equations). In case of absence of external
exchange similar equations should describe corresponding {\it internal}
exchange processes.
\vskip 0.2cm
Summerizing, we may assume the rule that the available changes of a
physical field system, or of a recognizable subsystem of a given system, {\it
must be referred somehow} to the very system, or to some of the susbsystems, in
order to evaluate their significance:

	-if the referred quantity is zero, then the changes are admissible and
the system/subsystem keeps its identity;

	-if mutually referred quantities among subsystems establish dynamical
equilibrium, i.e., each subsystem gains as much as it loses, then the whole
system keeps its identity;

	-if the referred quantity is not zero then the identity of the
system/subsystem is partially, or fully, lost, so, our system undergoes
essential changes leading to becoming subsystem of another system, or to
destruction, giving birth to new system(s).

 \vskip 0.3cm In trying to formalize these views it seems
appropriate to give some initial explicit formulations of some most basic
features (properties) of what we call physical object, which features would
lead us to a, more or less, adequate theoretical notion of our intuitive notion
of a physical object. Anyway, the following properties of the theoretical
concept "physical object" we consider as necessary:

      1. It can be created during finite period(s) of time.

      2. It can be destroyed during finite period(s) of time.

      3. It occupies finite 3-volume at any moment of its existence, so it has
spatial structure and may be considered as a system consisting of two or more
interconnected subsystems.

      4. It has a definite stability to withstand definite external
disturbances.

      5. It has definite conservation properties.

      6. It necessarily carries sufficiently universal measurable quantities,
e.g., energy-momentum.

      7. It exists in an appropriate environment (called usually vacuum), which
 provides all necessary existence needs. Figuratively speaking, every stable
physical object lives in a dynamical equilibrium with the outside world, which
dynamical equilibrium may be realized in various regimes.

      8. It can be detected by other physical objects through allowed
 exchanges of appropriate physical quantities, e.g., energy-momentum.

      9. It  may combine/coexist through interaction with other appropriate
physical objects to form new objects/systems of higher level structure. In
doing this it may keep its identity and can be recognized and identified
throughout the existence of the system as its constituent/subsystem.

      10. Its destruction gives necessarily birth to new objects, and this
process respects definite rules of conservation. In particular, the available
interaction energy among its subsystems may transform entirely, or partially, to
kinetic one, and carried away by the newly created objects/systems.

\vskip 0.3cm
The property {\it to be spatially finite} we consider as a very essential one.
So, the above features do NOT allow the classical material points and the
infinite classical fields (e.g.  plane waves) to be considered as physical
objects since the former have no structure and cannot be destroyed at all, and
the latter carry infinite energy, so they cannot be finite-time created. Hence,
the Born-Infeld "principle of finiteness" [2] stating that {\it "a satisfactory
theory should avoid letting physical quantities become infinite"} may be
strengthened as follows:
\vskip 0.3cm
{\bf All real physical
objects are spatially finite entities and NO infinite values of the physical
quantities carried by them are allowed}.

\vskip 0.2cm
Clearly, together with the purely qualitative features, physical objects carry
important quantitatively described physical properties, and any external
interaction may be considered as an exchange of the corresponding quantities
provided both the object and the corresponding environment carry them. Hence,
the more universal is a physical quantity the more useful for us it is, and
this moment determines the exclusively important role of energy-momentum, which
modern physics considers as the most universal one, i.e., it is more or less
assumed that: \vskip 0.3 cm

{\bf All physical objects necessarily carry energy-momentum and most of them
are able in a definite extent to lose and gain energy-momentum.}

\vskip 0.3 cm
The above notes more or less say that we make use of the term "physical object"
when we consider it from integral point of view, i.e. when its stability
against external perturbations is guaranteed. We make use of the term "physical
system" when time-stable interacting subsystems are possible to be
recognized/identified and the behavior of the system as a whole, i.e.,
considered from outside, we try to consider as seriously dependent on its
internal dynamical structure, i.e., on an available stable interaction of its
time-recognizable/time-identifiable subsystems. Therefore we shall follow the
rule:

\vskip 0.2cm
{\bf Physical recognizability of time-stable subsystems of a
physical system requires corresponding mathematical recognizability in the
theory}.
\vskip 0.2cm
We note that further we approach the problem from {\it nonrelativistic}
viewpoint.
\vskip 0.2cm
The above notes make us assume that any physical interaction presupposes
dynamical flows of some physical quantities among the subsystems of the
physical system considered. The field nature of the objects  suggests the {\it
local nature} of these flows, so, {\bf every continuous subsystem is supposed to
be able to build CORRESPONDING LOCAL INSTRUMENTS, realizing explicitly the
flows}. In static cases these flows reduce, of course, to stress. Formally this
means:
\begin{center}
\hfill\fbox{
    \begin{minipage}{0.97\textwidth}
\vskip 0.3cm
	1. We must have a mathematical field object $\mathcal{A}$ representing
the system as a whole.
\vskip 0.3cm
	2. The supposed existence of recognizable and mutually
interacting subsystems $(A_1,A_2,...)$ of $\mathcal{A}$ leads to the assumption
for {\it real} but {\it admissible}, i.e., not leading to annihilation, changes
of the subsystems, so, such changes should be formally represented by tensor
objects.
 \vskip 0.3cm
	3. The local flow manifestation of the admissible real changes suggests
to make use of appropriate combination of tensor objects, corresponding tensor
co-objects, and appropriate invariant differential operators.
 \vskip 0.3cm
	4. Every coupling inside this combination shall distinguish
existing partnership, i.e., interaction, among the subsystems, so, all such
couplings should be duly respected when the system's time-stability is to be
understood. 
\vskip 0.5cm \end{minipage}} \hfill \end{center}

We consider now how this can be realized if the physical appearance of a field
object can be formally represented by a vector field on the space
$\mathbb{R}^3$.

\section{Maxwell stress tensors generated by one and two vector fields}

Every vector field, defined on an arbitrary manifold $M$, generates 1-parameter
family $\varphi_t$ of (local in general) diffeomorphisms of $M$. Therefore,
having defined a vector field $X$ on $M$, we can consider for each
$t\in\mathbb{R}$ the corresponding diffeomorphic image $\varphi_t(U)$ of any
region $U\subset M$. Hence, interpreting the external parameter $t$ as time,
which is NOT obligatory, a vector field $X$ seems appropriate to be formally
tested as mathematical image of some spatially finite field object if:

	- the spatially finite object under consideration occupies for each
$t$ a finite region $U_t\subset\mathbb{R}^3$,

	- it propagates in the 3-space,

	- it stays recognizable and identifiable during propagation.

From physical point of view, however, the test of $X$ as mathematical image
of a physical object should acknowledge the abilities of the vector field to
build appropriate partner(s), as well as,
additional and appropriate mathematical images of such physically
important quantities like stress, energy and momentum, by means of appropriate
changes of which the object should realize every admissible interaction through
establishing and supporting corresponding balance relations.

 We are going to see further how vector fields on $\mathbb{R}^3$ meet such
challenges.

Let now $X$ be a vector field on the euclidean space $(\mathbb{R}^3,g)$, where
$g$ is the euclidean metric in $T\mathbb{R}^3$, having in the canonical global
coordinates $(x^1,x^2,x^3=x,y,z)$ components $g_{11}=g_{22}=g_{33}=1$ and
$g_{12}=g_{13}=g_{23}=0$. The induced euclidean metric in $T^*\mathbb{R}^3$ has
in the dual bases the same components and will be denoted further by the same
letter $g$. The corresponding isomorphisms between the tangent and cotangent
spaces and their tensor, exterior and symmetric products will be denoted by
the same simbol $\tilde{g}$, so (summation on the repeating indecies is assumed)
$$
\tilde{g}\left(\frac{\partial}{\partial x^i}\right)=
g^{ik}\left(\frac{\partial}{\partial x^k}\right)=dx^i,
\ \ (\tilde{g})^{-1}(dx^i)=\frac{\partial}{\partial x^i} \ \  \cdots
$$

Having a co-vector field, i.e. 1-form $\alpha$, or
another vector field  $Y$ on $\mathbb{R}^3$, we can form the flow of $X$
across $\alpha$, or across the $\tilde{g}$-coobject $\tilde{g}(Y)$ of $Y$:
$$
i_{X}\alpha=\langle\alpha,X\rangle=\alpha_1X^1+\alpha_2X^2+\alpha_3X^3,
$$
$$
i_{X}\tilde{g}(Y)=\langle\tilde{g}(Y),X\rangle=
g(X,Y)\equiv X.Y=g_{ij}X^iY^j=X_iY^i=X_1Y^1+X_2Y^2+X_3Y^3.
$$
This flow of $X$ is invariant entity, so to its admissible
and appropriate changes should be paid due respect. According to classical
vector analysis on $\mathbb{R}^3$ [16] for the differential of the function
$g(X,Y)$ we can write
$$
\mathbf{d}g(X,Y)=(X.\nabla)Y+(Y.\nabla)X+X\times\mathrm{rot}\,(Y)+
Y\times\mathrm{rot}\,(X),
$$
where in our coordinates
$$
X.\nabla=\nabla_{X}=X^i\frac{\partial}{\partial x^i}, \ \ \
(X.\nabla)Y=\nabla_{X}Y=
X^i\frac{\partial Y^j}{\partial x^i}\frac{\partial}{\partial x^j},
$$
"$\times$" is the usual vector product, and
$$
(\mathrm{rot}X)^i=\left(\frac{\partial X^3}{\partial x^2}-\frac{\partial
X^2}{\partial x^3}, \ \frac{\partial X^1}{\partial x^3}-\frac{\partial
X^3}{\partial x^1}, \ \frac{\partial X^2}{\partial x^1}-
\frac{\partial X^1}{\partial x^2}\right)\cdot
$$
The Hodge $*_{g}$-operator acts in these coordinates as follows:
$$
*dx=dy\wedge dz, \ \ *dy=-dx\wedge dz, \ \ *dz=dx\wedge dy,
$$
$$
*(dx\wedge dy)=dz, \ \ *(dx\wedge dz)=-dy, \ \ *(dy\wedge dz)=dx,
$$
$$
*(dx\wedge dy\wedge dz)=1, \ \ \ *1=dx\wedge dy\wedge dz .
$$
{\bf Corollary}. The following relation holds ($\mathbf{d}$ denotes the
exterior derivative):
$$
\mathrm{rot}X=(\tilde{g})^{-1}\,
*\,\mathbf{d}\,\tilde{g}(X), \ \ \text{or} \ \
\tilde{g}(\mathrm{rot}X)=*\,\mathbf{d}\,\tilde{g}(X).
$$
Assume now that in the above expression for $\mathbf{d}g(X,Y)$ we put $X=Y$,
i.e., we consider the invariant local change of the flow of $X$ across its
proper coobject $\tilde{g}(X)$. We obtain
$$
\frac12\mathbf{d}g(X,X)=\frac12\mathbf{d}(X^2)=
X\times\mathrm{rot}X+(X.\nabla)X=X\times\mathrm{rot}X+\nabla_XX.
$$
In components, the last term on the right reads
$$
(\nabla_XX)^j=X^i\nabla_i X^j=\nabla_i(X^iX^j)-X^j\nabla_iX^i=
\nabla_i(X^iX^j)-X^j\mathrm{div}\,X ,
$$
where (denoting by $L_X$ the Lie derivative along $X$)
$$
\mathrm{div}X=*L_X\left(dx\wedge dy\wedge dz\right)=
*\left(\frac{\partial X^i}{\partial x^i}dx\wedge dy\wedge dz\right)=
\frac{\partial X^i}{\partial x^i}.
$$
Substituting into the preceding relation, replacing
$\mathbf{d}(X^2)$ by $(\nabla_i\delta^i_jX^2)dx^j$, where $\delta^i_j$ is the
unit tensor in $T\mathbb{R}^3$, and making some elementary transformations we
obtain
$$
\nabla_i\left(X^iX^j-\frac12
g^{ij}X^2\right)= \big [(\mathrm{rot}\,X)\times X+X\mathrm{div}\,X\big ]^j.
$$
The symmetric 2-tensor
$$
M^{ij}=X^iX^j-\frac12\,g^{ij}X^2=
\frac12\Big
[X^iX^j+\big(\tilde{g}^{-1}*\tilde{g}(X)\big)^{ik}
\big(*\tilde{g}(X)\big)_{k}\,^j\Big ]
$$
we shall call further Maxwell stress tensor generated by the (arbitrary) vector
field $X\in\mathfrak{X}(\mathbb{R}^3)$. The components $M^i_j$ represent the
generated by the dynamical nature of $X$ local stresses, and the local stress
energy is represented in terms of $tr(M^i_j)=M^i_i$. Momentum is missing since
time is a missing dimension. {\bf The appropriate changes of $M^i_j$, given in
our case by $(rot\,X\times X)$ and $(X\,div\,X)$, should represent the local
instruments in terms of which the physical object, formally represented by $X$,
could establish balance relations with other objects}.

We specially note that, formally, $M^{ij}$ may be represented as sum of the
stresses carried by $X$ and by the 2-vector $\tilde{g}^{-1}*\tilde{g}(X)$.
Hence, since $\tilde{g}^{-1}*\tilde{g}(X)$ is uniquely determined by $X$ and
$g$, we may assume the viewpoint that the physical object considered is
formally represented by $X$ and $\tilde{g}^{-1}*\tilde{g}(X)$, moreover, to
consider $X$ and $\tilde{g}^{-1}*\tilde{g}(X)$ as images of two real
subsystems of our physical object. Finally we note that this view
excludes availability of interacting stress between these two subsystems:
$M^{ij}$ is sum of the stresses carried by $X$ and
$\tilde{g}^{-1}*\tilde{g}(X)$.

Clearly, when we raise and lower indices
in canonical coordinates with $\tilde{g}$ we shall have the following component
relations:
$$
M_{ij}=M_i^j=M^{ij}
$$ which does not mean, of course, that we equalize
quantities being elements of different linear spaces.

We note now the easily verified relation between the vector product "$\times$"
and the wedge product in the space of 1-forms on $\mathbb{R}^3$:
$$
X\times Y=(\tilde{g})^{-1}\,(*\,(\tilde{g}(X)\wedge\tilde{g}(Y)))
$$
$$
=(\tilde{g})^{-1}\,\circ i(X\wedge Y)(dx\wedge dy\wedge dz), \ \
X,Y\in\mathfrak{X}(\mathbb{R}^3) .
$$

We are going to consider now the differential flow nature of
$\nabla_iM^i_jdx^j$.

	{\bf Proposition}. If $\alpha=\tilde{g}(X)$ then the following relation
holds ($i(X)\mathbf{d}\alpha$ means $X^i\mathbf{d}\alpha_{ij}dx^j$):
$$
\tilde{g}(\mathrm{rot}\,X\times X)=i(X)\mathbf{d}\alpha=
-*(\alpha\wedge *\mathbf{d}\alpha) .
$$
{\it Proof}.
$$\tilde{g}(\mathrm{rot}\,X\times X)=
\tilde{g}\circ(\tilde{g})^{-1}\,*(\tilde{g}(\mathrm{rot}X)\wedge\tilde{g}(X))=
-*(\alpha\wedge *\mathbf{d}\alpha).
$$
For the component of $i(X)\mathbf{d}\alpha$ before $dx$ we obtain
$$
-X^2\left(\frac{\partial \alpha_2}{\partial x^1}-\frac{\partial
\alpha_1}{\partial x^2}\right)-X^3\left(\frac{\partial \alpha_3}{\partial
x^1}- \frac{\partial \alpha_1}{\partial x^3}\right),
$$
and the same quantity is easily obtained for the component of
$[-*(\alpha\wedge *\mathbf{d}\alpha)]$ before $dx$. The same is true for the
components of the two 1-forms before $dy$ and $dz$. The proposition is proved.

Hence, $\mathbf{d}\alpha=\mathbf{d}\tilde{g}(X)$ is the 2-form across which the
vector field $X$ will drag the points of the finite region
$U\subset\mathbb{R}^3$.

As for the second term
$X\mathrm{div}X$ of the divergence $\nabla_iM^{ij}$, since
$\mathbf{d}*\alpha=\mathrm{div}X(dx\wedge dy\wedge dz)$, we easily obtain
$$
i(\tilde{g}^{-1}(*\alpha))\mathbf{d}*\alpha=(\mathrm{div}X)\alpha.
$$
Hence, additionally,
the 2-vector $\tilde{g}^{-1}(*\alpha)$ will drag the points of $U\subset
\mathbb{R}^3$ across the 3-form $\mathbf{d}*\alpha$.

We can write now
$$
M_{ij}=\frac12[\alpha_i\alpha_j+(*\alpha)_i\,^k(*\alpha)_{kj}], \ \ \
\nabla_iM^i_j=\big[i(X)\mathbf{d}\alpha +
i(\tilde{g}^{-1}(*\alpha))\mathbf{d}*\alpha\big]_j .
$$
So, the stress balance $\nabla_iM^i_j=0$ is described by
$$
i(X)\mathbf{d}\alpha=-i(\tilde{g}^{-1}(*\alpha))\mathbf{d}*\alpha.
$$

One formal suggestion that comes from the above relations is that the interior
product of a (multi)vector and a differential form (i.e. the flow of a
(multi)vector field across a differential form) is appropriate quantity to be
used as a quantitative measure of local physical interaction.

Another suggestion is that our physical field object, initially represented by
the vector field $X$, can be equally well represented by
$\alpha=\tilde{g}(X)$ and $*\alpha$. Viewed this way, we also see that
{\it there is no available local nonzero interaction stress
between $\alpha$ and $*\alpha$}.

Hence, the naturally isolated two
terms in $\nabla_iM^i_jdx^j$ suggest: any realizable dynamical stress,
represented by the vector field $X$, to be described by
$(\alpha,*\alpha)$, and the time-recognizable nature of $\alpha$ and $*\alpha$
to be guaranteed by the balance equation $i(X)\mathbf{d}\alpha=-
i(\tilde{g}^{-1}(*\alpha))\mathbf{d}*\alpha$,
or by
$$
i(X)\mathbf{d}\alpha=0, \ \
i(\tilde{g}^{-1}(*\alpha))\mathbf{d}*\alpha=0, \ \ \text{i.e.} \ \
X\times\mathrm{rot}X=0, \ \mathrm{div}X=0.
$$
Recalling now how the Lie derivative acts on 1-forms and 2-forms [17], namely,
$$
L_X\alpha=\mathbf{d}\langle\alpha,X\rangle+i(X)\mathbf{d}\alpha, \ \
\text{i.e.} \ \ \ \
L_X\alpha-\mathbf{d}\langle\alpha,X\rangle=i(X)\mathbf{d}\alpha,
$$
$$
L_{\bar{*\alpha}}*\alpha=\mathbf{d}\langle*\alpha,\bar{*\alpha}\rangle-
(-1)^{deg{(\bar{*\alpha})}.deg{(\mathbf{d})}}i(\bar{*\alpha})\mathbf{d}*\alpha,
\ \ \text{i.e.} \ \ \
L_{\bar{*\alpha}}*\alpha-\mathbf{d}\langle*\alpha,\bar{*\alpha}\rangle=
-i(\bar{*\alpha})\mathbf{d}*\alpha, \ \
$$
where $deg(\bar{*\alpha})=2, \  deg{(\mathbf{d})}=1$,
we see that the flow of $X$ across the 2-form $\mathbf{d}\alpha$, and the flow
of $\tilde{g}^{-1}(*\alpha)$ across $\mathbf{d}*\alpha$, are given by the
difference between two well defined coordinate free quantities, and this
difference determines when the local change of $\alpha$, resp. $*\alpha$, with
respect to $X$, resp. $\tilde{g}^{-1}(*\alpha)$, cannot be represented by
$\mathbf{d}\langle\alpha,X\rangle$,
resp. $\mathbf{d}\langle*\alpha,\tilde{g}^{-1}(*\alpha)\rangle$.

In the above consideration the role of the euclidean metric $g$ was somehow
overlooked since no derivatives of $g$ appeared explicitly. We are going now to
come to these equations paying due respect to $g$ .

We shall work further  with the 1-form
$\alpha=\tilde{g}(X)$, where the vector field $X$
generates stress according to
the Maxwell stress tensor $M^i_j(X)$. We want to see how the 3-form
$\alpha\wedge *\alpha=i(X)\alpha.dx\wedge dy\wedge dz=i(X)\alpha.\omega$
changes along an arbitrary vector
field $Y\neq X$, so we have to find the corresponding Lie derivative.
$$
L_Y(\alpha\wedge *\alpha)=(L_Y\alpha)\wedge *\alpha+\alpha\wedge L_Y(*\alpha)
$$
$$ = (L_Y\alpha)\wedge *\alpha+\alpha\wedge
*L_Y\alpha+\alpha\wedge[L_Y,*_1]\alpha= 2(L_Y\alpha)\wedge
*\alpha+\alpha\wedge[L_Y,*_1]\alpha  ,
$$
where $[L_Y,*_1]$ is the commutetor
$L_Y\circ\,*_1-*_1\circ L_Y$, and the index of $*$ denotes the degree of the
form it is applied to. Further we get
$$
(L_Y\alpha)\wedge
*\alpha=(i_Y\mathbf{d}\alpha)\wedge*\alpha-\langle\alpha,Y\rangle.\mathbf{d}*\alpha+
\mathbf{d}(\langle\alpha,Y\rangle.*\alpha).
$$
Noting that $L_Y(\alpha\wedge *\alpha)=\mathbf{d}(\alpha^2i_Y\omega)$
and the relation between the exterior derivative
$\mathbf{d}$ and the coderivative $\delta=(-1)^{3p+3+1}*\,\mathbf{d}\,*$
in euclidean case for p-forms, given by
$$
(-1)^{p(n-p)}\delta_p\circ *_{n-p}=*_{(n-p+1)}\circ \mathbf{d}_{(n-p)} ,
$$
we obtain consecutively
$$
\mathbf{d}\left(\frac12\alpha_i\alpha^ii_Y\omega-
\langle\alpha,Y\rangle*\alpha\right)=
-[\alpha^i(\mathbf{d}\alpha)_{ij}+\alpha_j\,\mathrm{div}X]Y^j\omega
+\frac12\alpha\wedge[L_Y,*_1]\alpha ,
$$
$$
\delta\circ
*_2\left(\frac12\alpha_i\alpha^i\,i_Y\omega-\langle\alpha,Y\rangle*\alpha\right)=
-[\alpha^i(\mathbf{d}\alpha)_{ij}+\alpha_j\,\mathrm{div}X]Y^j
+\frac12\alpha\wedge[L_Y,*_1]\alpha ,
$$
$$
*_2\langle\alpha,Y\rangle*\alpha=\langle\alpha,Y\rangle.\alpha,
$$
$$
\delta\left(\frac12\alpha_i\alpha^ig_{jk}-\alpha_j\alpha_k)Y^jdx^k\right)=
-\nabla_j\left[\left(\frac12\alpha_i\alpha^i\delta^j_k-
\alpha_k\alpha^j\right)Y^k\right] .
$$
So we get
$$
\nabla_j\left[\left(\frac12\alpha_i\alpha^i\delta^j_k-
\alpha_k\alpha^j\right)Y^k\right]=
[\alpha^i(\mathbf{d}\alpha)_{ij}+\alpha_j\,\mathrm{div}X]Y^j+
*\frac12\alpha\wedge[L_Y,*_1]\alpha .
$$
Hence, if $[L_Y,*_1]=0$, for example when $Y$ is a Killing vector field for $g$,
and the equations for our 1-form $\alpha=\tilde{g}(X)$ are
$$
\alpha^i(\mathbf{d}\alpha)_{ij}=0, \ \ \alpha_j\,\mathrm{div}X=0 , \ \
i,j=1,2,3 ,
$$
then the co-closed 1-form $M_{ij}Y^idx^j$ defines the closed
2-form $*(M_{ij}Y^idx^j)$, so, its integral over a closed 2-surface that
separates the 3-volume where, by supposition, $\alpha$ is different from zero,
gives a conservative quantity and the nature of this conservative quantity is
connected with the nature of the vector field $Y$, which is a Killing vector in
our 3-dimensional case. Of course, the term "conservative" here is, more or
less, trivial, since just stress is available, no space propagation takes
place.

As for solutions of the above equations, we may suppose two classes of
solutions:

	1. Linear, i.e. those satisfying $\mathbf{d}\alpha=0,
\mathbf{d}*\alpha=0$. These are generated by a function $f$ satisfying the
Laplace equation $\Delta f=0$  with $\alpha=\mathbf{d}f$ .

	2. Those, satisfying
$\mathbf{d}\alpha\neq 0$ but $\alpha^i\mathbf{d}\alpha_{ij}=0$ and
$\mathbf{d}*\alpha=0$. Note that for such nontrivial solutions the determinant
$det||(\mathbf{d}\alpha)_{ij}||, i,j=1,2,3$, is always zero, so the
homogenious linear system
$$
\alpha^i(x,y,z)\mathbf{d}\alpha_{ij}(x,y,z)=0
$$
allows to express explicitly $X$ through the derivatives of its components.

Of course, all these solutions are {\it static}, so they can NOT serve as
models of spatially propagating finite physical objects with dynamical
structure. \vskip 0.3cm

We pass now to the case of {\it two} vector fields.
\vskip 0.2cm
Let $V$ and $W$ be two vector fields on our euclidean 3-space. Summing up the
corresponding two Maxwell stress tensors we obtain the identity:

\setlength\arraycolsep{8pt}
\begin{eqnarray*}
\lefteqn{ \nabla_iM^{ij}_{(V,W)}\equiv \nabla_i\left(V^iV^j+W^iW^j-
\delta^{ij}\frac{V^2+W^2}{2}\right)={} } \nonumber\\ & & {}=\big
[(\mathrm{rot}\,V)\times V+ V\mathrm{div}\,V+
(\mathrm{rot}\,W)\times W+W\mathrm{div}\,W\big ]^j.
\end{eqnarray*}
Note that the balance in this case may look like, for example, as follows:
$$
(\mathrm{rot}\,V)\times V=-W\mathrm{div}\,W, \ \ \
(\mathrm{rot}\,W)\times W=-V\mathrm{div}\,V,
$$
which suggests internal stress coupling between $V$ and $W$.

Let now $(a(x,y,z),b(x,y,z))$ be two arbitrary functions on $\mathbb{R}^3$. We
consider the transformation
$$
(V,W)\rightarrow (V\,a-W\,b,V\,b+W\,a).
$$

{\bf Corollary.}

The tensor $M_{(V,W)}$ transforms to $(a^2+b^2)M_{(V,W)}$.

	{\bf Corollary.}

The transformations
$(V,W)\rightarrow (V\,a-W\,b,V\,b+W\,a)$ do not change the eigen
directions structure of $M^{ij}_{(V,W)}$.

	{\bf Corollary.}

If $a=\mathrm{cos}\,\theta, b=\mathrm{sin}\,\theta$, where
$\theta=\theta(x,y,z)$ then the tensor $M_{(V,W)}$ stays invariant:
$$
M_{(V,W)}=M_{(V\mathrm{cos}\,\theta-W\mathrm{sin}\,\theta,
V\mathrm{sin}\,\theta+W\mathrm{cos}\,\theta)}.
$$

The expression inside the parenteses above, denoted by $M^{ij}_{(V,W)}$, looks
formally the same as the introduced by Maxwell tensor
$M^{ij}(\mathbf{E},\mathbf{B})$ from physical considerations concerned with the
electromagnetic stress energy properties of continuous media in presence of
external electromagnetic field $(\mathbf{E},\mathbf{B})$. Hence, any vector
$V$, or any couple of vectors $(V,W)$, defines such tensor which we denote by
$M_V$, or $M_{(V,W)}$, and call {\bf Maxwell stress tensor}. The term, "stress"
in this general mathematical setting could be justified by the above mentioned
dynamical nature of vector fields. It deserves noting here that the two-vector
case  should be expected to satisfy some conditions of compatability between
$V$ and $W$ in order to represent a physical time stable stress flows.

We emphasize the following moments:
\vskip 0.2cm
	{\bf 1}. The differential identity satisfied by $M_{(V,W)}$
is purely mathematical;

	{\bf 2}. On the two sides of this identity stay well defined
coordinate free quantities;

	{\bf 3}. The tensors $M_{(V,W)}$ do not introduce
interaction stress: the full stress is the sum of the stresses generated by
each one of the couple $(V,W)$.
\vskip 0.2cm
Physically, we may say that the corresponding physical medium that occupies the
spatial region $\mathbf{U}_o$ and is parametrized by the points of the
mathematical subregion $U_o\subset\mathbb{R}^3$, is subject to {\it compatible}
and {\it admissible} physical "stresses", and these physical stresses are
quantitatively described by the corresponding physical interpretation of the
tensor $M_{(V,W)}$. Clearly, we could extend the couple $(V,W)$ to more vectors
$(V_1,V_2,...,V_p)$, but then the mentioned invariance properties of
$M_{(V,W)}$ may be lost, or should be appropriately extended.

We note that the stress tensor $M^{ij}$ appears as been subject to the
divergence operator, and if we interpret the components of $M^{ij}$ as physical
stresses, then its divergence acquires, in general, the physical interpretation
of force density. Of course, in the static situation as it is given by the
relation considered, no stress propagation is possible, so at every point the
local forces mutually compensate: $\nabla_{i}M^{ij}=0$. In case of
electromagnetic pure field objects where space propagation is {\bf necessary},
then the force field may NOT be zero: $\nabla_{i}M^{ij}\neq 0$, and we should
identify $\nabla_{i}M^{ij}\neq 0$ as a {\bf real time-change} of appropriately
defined momentum density $\mathbf{P}_{(\mathbf{E},\mathbf{B})}$. So, assuming
some expression for this {\bf time-dependent} momentum density
$\mathbf{P}_{(\mathbf{E},\mathbf{B})}$ we can understand the {\it dynamical
appearance} of our object through equalizing the spatially directed
uncompensated force densities $\nabla_{i}M^{ij}$ with the momentum density
changes along the time variable, i.e., equalizing $\nabla_iM^{ij}$ with the
$(ct)$-derivative of $\mathbf{P}_{(\mathbf{E},\mathbf{B})}$, where $c=const$ is
the translational propagation velocity of the momentum density flow of the
physical object/system considered. In order to find how to choose
$\mathbf{P}_{(\mathbf{E},\mathbf{B})}$ in case of {\it free} EM-field we have
to turn to the intrinsic {\it physical} properties of the field. As the past
years showed, namely $M^{ij}_{(\mathbf{E},\mathbf{B})}$ is the appropriate
carrier of the physical properties of the field, so, it seems natural to turn
to the eigen properties of $M^{ij}_{(\mathbf{E},\mathbf{B})}$.

\section{Eigen properties of Maxwell stress tensor}

We consider $M^{ij}(\mathbf{E},\mathbf{B})$ at some point $p\in\mathbb{R}^3$
and assume that in general the vector fields $\mathbf{E}$ and $\mathbf{B}$ are
lineary independent, so $\mathbf{E}\times\mathbf{B}\neq 0$. Let the coordinate
system be chosen such that the coordinate plane $(x,y)$ to coincide with the
plane defined by $\mathbf{E}(p),\mathbf{B}(p)$. In this coordinate system
$\mathbf{E}=(E_1,E_2,0)$ and $\mathbf{B}=(B_1,B_2,0)$, so, identifying the
contravariant and covariant indices through the Euclidean metric $\delta^{ij}$
(so that $M^{ij}=M^i_j=M_{ij}$), we obtain the following nonzero components of
the stress tensor: $$ M^1_1=(E^1)^2+(B^1)^2-\frac12(\mathbf{E}^2+\mathbf{B}^2);
\ \ M^1_2=M^2_1=E^1\,E_2+B_1\,B^2; $$ $$
M^2_2=(E^2)^2+(B^2)^2-\frac12(\mathbf{E}^2+\mathbf{B}^2); \ \
M^3_3=-\frac12(\mathbf{E}^2+\mathbf{B}^2). $$ Since $M^1_1=-M^2_2$, the trace
of $M$ is $Tr(M)=-\frac12(\mathbf{E}^2+\mathbf{B}^2)$.

The eigen value equation acquires the simple
form
$$
\big[(M^1_1)^2-(\lambda)^2\big]+(M^1_2)^2\big](M^3_3-\lambda)=0.
$$
The corresponding eigen values are
$$
\lambda_1=-\frac12(\mathbf{E}^2+\mathbf{B}^2);\ \
\lambda_{2,3}=\pm\sqrt{(M^1_1)^2+(M^1_2)^2}= \pm\frac12\sqrt{(I_1)^2+(I_2)^2} ,
$$
where
$I_1=\mathbf{B}^2-\mathbf{E}^2,\, I_2=2\mathbf{E}.\mathbf{B}$.

The corresponding to
$\lambda_1$ eigen vector $Z_1$ must satisfy the equation
$$
\mathbf{E}(\mathbf{E}.Z_1)+\mathbf{B}(\mathbf{B}.Z_1)=0,
$$
 and since
$(\mathbf{E},\mathbf{B})$ are lineary independent, the two coefficients
$(\mathbf{E}.Z_1)$ and $(\mathbf{B}.Z_1)$ must be equal to
zero, therefore, $Z_1\neq 0$ must be orthogonal to $\mathbf{E}$ and
$\mathbf{B}$, i.e. $Z_1$ must be colinear to $\mathbf{E}\times\mathbf{B}$:

The other two eigen vectors $Z_{2,3}$
satisfy correspondingly the equations
$$ \mathbf{E}(\mathbf{E}.Z_{2,3})+
\mathbf{B}(\mathbf{B}.Z_{2,3})=\Big[\pm\frac12\sqrt{(I_1)^2+(I_2)^2}+
\frac12(\mathbf{E}^2+\mathbf{B}^2)\Big]Z_{2,3}.         \ \ \ \ \ \    (*)
$$
Taking into account the easily verified relation
$$
\frac14\Big[(I_1)^2+(I_2)^2\Big]=
\left(\frac{\mathbf{E}^2+\mathbf{B}^2}{2}\right)^2-
|\mathbf{E}\times\mathbf{B}|^2 ,
$$
so that
$$
\frac{\mathbf{E}^2+\mathbf{B}^2}{2}- |\mathbf{E}\times\mathbf{B}|\geq 0 \ ,
$$
we conclude that the coefficient before $Z_{2,3}$ on the right is always different
from zero, therefore, the eigen vectors $Z_{2,3}(p)$ lie in the plane defined
by $(\mathbf{E}(p),\mathbf{B}(p)), \ p\in \mathbb{R}^3$. In particular,
the above mentioned transformation properties of the
Maxwell stress tensor $M(V,W)\rightarrow (a^2+b^2)M(V,W)$ show that the
corresponding eigen directions do not change under the transformation
$(V,W)\rightarrow (V\,a-W\,b,V\,b+W\,a)$.
\vskip 0.2cm
The above consideration suggests: {\it the intrinsically allowed
dynamical abilities of the field are: translational
along $(\mathbf{E}\times\mathbf{B})$, and rotational inside the plane defined
by $(\mathbf{E},\mathbf{B})$, hence, we may expect finding field objects the
propagation of which shows intrinsic local consistency between rotation and
translation.}
\vskip 0.2cm
It is natural to ask now under what conditions the very $\mathbf{E}$ and
$\mathbf{B}$ may be eigen vectors of $M(\mathbf{E},\mathbf{B})$? Assuming
$\lambda_2=\frac12\sqrt{(I_1)^2+(I_2)^2}$ and $Z_2=\mathbf{E}$
in the above relation and having in view that
$\mathbf{E}\times\mathbf{B}\neq 0$ we obtain that
$\mathbf{E}(\mathbf{E}^2)+\mathbf{B}(\mathbf{E}.\mathbf{B})$ must be
proportional to $\mathbf{E}$, so, $\mathbf{E}.\mathbf{B}=0$,
i.e. $I_2=0$. Moreover, substituting now $I_2=0$ in that same  relation we obtain
$$
\mathbf{E}^2=\frac12(\mathbf{B}^2-\mathbf{E}^2)+
\frac12(\mathbf{E}^2+\mathbf{B}^2)=\mathbf{B}^2, \ \ \text{i.e.}, \ \ I_1=0.
$$
The case
"-" sign before the square root, i.e. $\lambda_3=-\frac12\sqrt{(I_1)^2+(I_2)^2}$,
leads to analogical conclusions just the role of $\mathbf{E}$ and $\mathbf{B}$
is exchanged.

\vskip 0.2cm
{\bf Corollary}. $\mathbf{E}$ and $\mathbf{B}$ may be eigen
vectors of $M(\mathbf{E},\mathbf{B})$ only if
 $I_1=I_2=0$.
\vskip 0.2cm

The above notices suggest to consider in a more detail the case
$\lambda_2=-\lambda_3=0$ for the vacuum case. We shall show, making use of the
Lorentz transformation in 3-dimensional form that, if
these two relations do not hold then under $\mathbf{E}\times\mathbf{B}\neq 0$
the translational velocity of propagation
is less then the speed of light in vacuum $c$. Recall first the
transformation laws of the electric and magnetic vectors under Lorentz
transformation defined by the 3-velocity vector $\mathbf{v}$ and corresponding
parameter $\beta=v/c, v=|\mathbf{v}|$. If $\gamma$ denotes the factor
$1/\sqrt{1-\beta^2}$ then we have $$
\mathbf{E'}=\gamma\,\mathbf{E}+\frac{1-\gamma}{v^2}\mathbf{v}(\mathbf{E}.
\mathbf{v})+\frac{\gamma}{c}\mathbf{v}\times\mathbf{B} , $$ $$
\mathbf{B'}=\gamma\,\mathbf{B}+\frac{1-\gamma}{v^2}\mathbf{v}(\mathbf{B}.
\mathbf{v})-\frac{\gamma}{c}\mathbf{v}\times\mathbf{E} . $$

Assume first that $I_2=2\mathbf{E}.\mathbf{B}=0$, i.e. $\mathbf{E}$ and $\mathbf{B}$ are
orthogonal, so, in general, in some coordinate system we shall have
$\mathbf{E}\times\mathbf{B}\neq 0$ .

If $I_1>0$, i.e. $|\mathbf{E}|<|\mathbf{B}|$, we shall show that the
conditions $\mathbf{E'}=0, \mathbf{v}.\mathbf{B}=0, \infty>\gamma>0$ are
compatible. In fact, these assumptions lead to
$\gamma\,\mathbf{v}.\mathbf{E}+(1-\gamma)(\mathbf{E}.\mathbf{v})=0$, i.e.
$\mathbf{E}.\mathbf{v}=0$. Thus,
 $c|\mathbf{E}|=v|\mathbf{B}||\mathrm{sin}(\mathbf{v},\mathbf{B})|$, and since
$\mathbf{v}.\mathbf{B}=0$ then $|\mathrm{sin}(\mathbf{v},\mathbf{B})|=1$.
It follows that the speed
$v=c\frac{|\mathbf{E}|}{|\mathbf{B}|}<c$ is allowed.

If $I_1<0$, i.e. $|\mathbf{E}|>|\mathbf{B}|$, then the conditions
$\mathbf{B'}=0$ and $\mathbf{v}.\mathbf{E}=0$ analogically lead to the
conclusion that the speed $v=c\frac{|\mathbf{B}|}{|\mathbf{E}|}<c$ is allowed.

Assume now that $I_2=2\mathbf{E}.\mathbf{B}\neq 0$. We are looking for a reference frame
$K'$ such that $\mathbf{E'}\times\mathbf{B'}=0$, while in the reference frame $K$ we have
$\mathbf{E}\times\mathbf{B}\neq 0$. We choose the relative velocity $\mathbf{v}$ such
that $\mathbf{v}.\mathbf{E}= \mathbf{v}.\mathbf{B}=0$. Under these conditions the
equation $\mathbf{E'}\times\mathbf{B'}=0$ reduces to $$
\mathbf{E}\times\mathbf{B}+\frac{\mathbf{v}}{c}(\mathbf{E}^2+\mathbf{B}^2)=0 ,\ \
\text{so}, \ \ \frac{v}{c}=|\mathbf{E}\times\mathbf{B}|/(\mathbf{E}^2+\mathbf{B}^2). $$
Now, from the above mentioned inequality
$\mathbf{E}^2+\mathbf{B}^2-2|\mathbf{E}\times\mathbf{B}|\geq 0$ it follows that
$\frac{v}{c}<1$.

Physically, these considerations show that under nonzero $I_1$ and $I_2$ the
translational velocity of propagation of the field, and of the stress field
energy density of course, can be eliminated by passing to appropriate frame of
mass bodies. Hence, the only realistic choice for the vacuum case, where this
velocity is assumed by definition to be equal to the vacuum light speed $c$
and can not be eliminated by passing to appropriate physical frame, is
$I_1=I_2=0$, which is equivalent to
$\mathbf{E}^2+\mathbf{B}^2=2|\mathbf{E}\times\mathbf{B}|$. In view of this and
assuming $|Tr(M)|=\frac12(\mathbf{E}^2+\mathbf{B}^2)$ to be the stress energy
density of the field, the names "electromagnetic energy flux" for the quantity
$c\mathbf{E}\times\mathbf{B}$, and "momentum" for the quantity
$\frac1c\mathbf{E}\times\mathbf{B}$, seem well justified without turning to any
dynamical field equations.

These considerations suggest also that if $I_1=0$, i.e.
$|\mathbf{E}|^2=|\mathbf{B}|^2$ during propagation, then the energy density can
be presented in terms of each of the two constituents, moreover, in this
respect, both constituents have the same rights. Therefore, a local mutual
energy exchange between the supposed two subsystems formally represented
by appropriate combinations of $\mathbf{E},\mathbf{B}$ is not forbidden in
general, but, if it takes place, it must be {\it simultaneous} and in {\it
equal quantities}. Hence, under zero invariants $I_1=0$ and
$I_2=2\mathbf{E}.\mathbf{B}=0$, internal energy redistribution between the two
subsystems of the field would be allowed, but such an exchange should occur
{\it without available interaction energy}.

The following question now arizes: is it {\it physically} allowed to interprit
each of the two vector fields  $\mathbf{E},\mathbf{B}$ as mathematical image of
a recognizable time-stable physical subsystem of the EM-field?

Trying to answer this question we note that the relation
$\mathbf{E}^2+\mathbf{B}^2=2|\mathbf{E}\times\mathbf{B}|$ and the required
time-recognizability during propagation (with velocity "c") of each subsystem
of the field suggest/imply also that {\it each of the two subsystems must be
able to carry locally momentum and to exchange locally momentum with the other
one}, since this relation means that the energy density is always strongly
proportional to the momentum density magnitude
$\frac1c|\mathbf{E}\times\mathbf{B}|$. Hence, the couple
$(\mathbf{E},\mathbf{B})$ is able to carry momentum, but neither of
$\mathbf{E},\mathbf{B}$ can carry momentum separately. Moreover, the important
observation here is that, verious combinations constructed out of the
constituents $\mathbf{E}$ and $\mathbf{B}$, e.g.,
$(\mathbf{E}\,cos\theta-\mathbf{B}\,sin\theta,
\mathbf{E}\,sin\theta+\mathbf{B}\,cos\theta)$, where $\theta(x,y,z;t)$ is a
functon, may be considered as possible representatives of the two recognizable
subsystems since they carry the same energy
$\frac12(\mathbf{E}^2+\mathbf{B}^2)$ and momentum
$\frac1c|\mathbf{E}\times\mathbf{B}|$ densities. Therefore, the suggestion by
Maxwell vacuum equations that the very $\mathbf{E}$ and $\mathbf{B}$ may be
considered as appropriate mathematical images of recognizable time-stable
subsystems of a time-dependent and space propagating electromagnetic field
object does NOT seem adequate and has to be reconsidered.

Hence, which combinations of $\mathbf{E}$ and $\mathbf{B}$ deserve to represent
mathematically the two subsystems of a time-dependent and space-propagating
electromagnetic field object?

In view of these considerations we assume the following understanding:
\vskip 0.3cm
{\bf Every real
EM-field is built of two recognizable subsystems, the mathematical images of
which are not the very $(\mathbf{E},\mathbf{B})$, but are
expressed in terms of $(\mathbf{E},\mathbf{B})$, both these
subsystems carry always the same quantity of energy-momentum, guaranteeing in
this way that the supposed internal energy-momentum exchange will also be
in equal quantities and simultanious}.

\section{Real EM-fields viewed as built of two recognizable and permanently
interacting subsystems.}

In accordance with the above assumption the description of dynamical and
space-propagating behavior of the field will need two appropriate mathematical
objects to be constructed out of the two constituents
$(\mathbf{E},\mathbf{B})$. These two mathematical objects must meet the
required property that the two physical subsystems of the field carry always
the same quantity of energy-momentum, and that any possible internal
energy-momentum exchange between the two subsystems shall be {\it simultaneous}
and {\it in equal quantities}.

We are going to consider time dependent fields, and begin with noting once
again the assumption that the full stress tensor (and the energy density, in
particular) is a sum of the stress tensors carried separately by the two
subsystems. As we mentiond above, this does NOT mean that there is no energy
exchange between the two subsystems of the field.

Now, following the above stated idea we have to find two appropriate
mathematical images of the field which images are NOT  represented directly by
the electric $\mathbf{E}$ and magnetic $\mathbf{B}$ vectors, but are
constructed out of them. In terms of these two appropriate mathematical
representatives of the corresponding two partnering subsystems we must express
the mentioned  special kind of energy-momentum exchange, respecting in this way
the fact that {\it neither} of the two constituents
$(\mathbf{E},\mathbf{B})$ {\it is able to carry momentum separately}.

In view of the above we have to assume that the field keeps its identity through
adopting some special and appropriate dynamical behavior according to its
intrinsic capabilities. Since the corresponding dynamical/field behavior must
be consistent with the properties of the intrinsic stress-energy-momentum nature
of the field, we come to the conclusion that Maxwell stress tensor
$M(\mathbf{E},\mathbf{B})$ should play the basic role, and its zero-divergence
in the static case should suggest how to determine the appropriate
structure and allowed dynamical propagation.

Recall that any member of the family
$$
(\mathcal{E},\mathcal{B})=(\mathbf{E},\mathbf{B},\theta)=
(\mathbf{E}\,\mathrm{cos}\,\theta-
\mathbf{B}\,\mathrm{sin}\,\theta; \ \mathbf{E}\,\mathrm{sin}\,\theta+
\mathbf{B}\,\mathrm{cos}\,\theta), \ \ \theta=\theta(x,y,z;t),
$$
generates the same Maxwell stress tensor. So, the most natural assumption should
read like this:

{\bf Any member
$(\mathbf{E},\mathbf{B},\theta_1)$
of this $\theta$-family is looking for an energy-momentum exchanging partner
$(\mathbf{E},\mathbf{B},\theta_2)$ inside the family, and identifies itself
through appropriate (local) interaction with the partner found, defining in
this way corresponding dynamical behavior of the field}.

Simply speaking, a time-dependent EM-field is formally represented by two
members of the above $\theta$-family, and the coupling
$(\mathbf{E},\mathbf{B},\theta_1)
\leftrightarrow(\mathbf{E},\mathbf{B},\theta_2)$ is unique.


Note that working with $\alpha$-invariant quantities, e.g. $M^{ij}$ and
$\mathbf{E}\times\mathbf{B}$, we may consider the couple
$(\mathbf{E},\mathbf{B})$ as any member of the $\alpha$-family. In view of this
we shall make use of the local divergence of the Maxwell stress tensor and the
time derivative of the local momentum flow of the field in order to find
the corresponding partner-subsystem of $(\mathbf{E},\mathbf{B})$.

Further we shall call these two subsystems just partner-fields.

Recall the divergence \setlength\arraycolsep{8pt}
\begin{eqnarray*} \lefteqn{ \nabla_iM^{ij}\equiv
\nabla_i\left(\mathbf{E}^i\mathbf{E}^j+\mathbf{B}^i\mathbf{B}^j-
\delta^{ij}\frac{\mathbf{E}^2+\mathbf{B}^2}{2}\right)={} } \nonumber\\ & & {}= \big
[(\mathrm{rot}\,\mathbf{E})\times \mathbf{E}+ \mathbf{E}\mathrm{div}\,\mathbf{E}+
(\mathrm{rot}\,\mathbf{B})\times \mathbf{B}+ \mathbf{B}\mathrm{div}\,\mathbf{B}\big ]^j.
\end{eqnarray*}

As we mentioned, in the static case, i.e. when the vector fields
$(\mathbf{E},\mathbf{B})$ do not depend on the time "coordinate" $\xi=ct$, NO
propagation of field momentum density $\mathbf{P}$ should take place, so, at
every point, where $(\mathbf{E},\mathbf{B})\neq 0$, the  stress generated
forces must mutually compensate, i.e. the divergence $\nabla_iM^{ij}$ should be
equal to zero: $\nabla_iM^{ij}=0$. In this static case Maxwell vacuum equations
$$
\mathrm{rot}\,\mathbf{E}+\frac{\partial\mathbf{B}}{\partial \xi}=0,\quad
\mathrm{rot}\,\mathbf{B}-\frac{\partial\mathbf{E}}{\partial \xi}=0,\quad
\mathrm{div}\,\mathbf{E}=0,\quad \mathrm{div}\,\mathbf{B}=0 \ \ \
(\text{J.C.M.})
$$
give: $\mathrm{rot}\mathbf{E}=\mathrm{rot}\mathbf{B}=0;\,
\mathrm{div}\mathbf{E}=\mathrm{div}\mathbf{B}=0$, so, all static solutions to
Maxwell equations determine a sufficient, but NOT necessary, condition that
brings to zero the right hand side of the divergence through forcing each of
the four vectors there to get zero values.

In the non-static case, i.e. when $\frac{\partial\mathbf{E}}{\partial t}\neq 0;
\,\frac{\partial\mathbf{B}}{\partial t}\neq 0$, time change and propagation of
field momentum density should be expected, so, a full mutual compensation of the
generated by the Maxwell stresses at every spatial point local forces may NOT
be possible, which means $\nabla_iM^{ij}\neq 0$ in general. These local forces
generate time-dependent momentum inside the corresponding region.
Therefore, if we want to describe this physical process of field
energy-momentum density time change and spatial propagation we have to
introduce explicitly the dependence of the local momentum vector
field $\mathbf{P}$ on $(\mathbf{E},\mathbf{B})$, and to express the flow of the
electromagnetic energy-momentum in terms of the available stress-flows.
The spatial aspects of the flow-process are represented by the two
terms of $\nabla_iM^{ij}\neq 0$, and the time aspects of the process should come
from the time derivative of $\mathbf{P}$, considered as
a function of the two constituents $(\mathbf{E},\mathbf{B})$. We show now how,
following the classical formal approach, we can come to the desired relation.

Note that compare to classical approach where the flows of
the very $\mathbf{E},\mathbf{B}$ through some 2-surface $S$ are considered, we
consider flows of quantities having direct stress-energy-momentum change sense,
since these physical quantities determine the dynamical appearance of our
EM-object.

In terms of $\mathfrak{F}^j=\nabla_iM^{ij}\neq 0$ the 2-form that is to be
integrated on $S$ is given by reducing
$*\tilde{g}(\mathfrak{F})$ on $S$,
where $*$ denotes the euclidean Hodge $*$. On the other hand, the local
momentum density flow time change across $S$ should naturally be represented by
$\frac{d}{dt}\int_{S}*\tilde{g}(\mathbf{P}(\mathbf{E},\mathbf{B}))$.
Recalling that $t$ is considered as external parameter and passing to
corresponding integrals on $S$ we can write:
$$
\frac{d}{dt}\int_{S}*\tilde{g}(\mathbf{P(\mathbf{E},\mathbf{B})})=
\int_{S}*\tilde{g}(\mathfrak{F}).
$$
 \vskip 0.3cm
 The explicit expression for
$\mathbf{P}(\mathbf{E},\mathbf{B})$, paying due respect to J.Poynting, and
to J.J.Thomson, H.Poincare, M. Abraham, and in view of the huge, a century
and a half available experience, has to be introduced by the following
\vskip
0.4cm \noindent {\bf Assumption}: {\it The entire field momentum density is
given by $\mathbf{P}:=\frac1c\mathbf{E}\times\mathbf{B}$} .
 \vskip 0.4cm
According to this {\bf Assumption}, to the above interpretation of the relation
$\nabla_iM^{ij}\neq 0$, in view of the assumed by us local
energy-momentum exchange approach to description of the dynamics of the field,
in vector field terms and in canonical coordinates on $\mathbb{R}^3$ we come
the following vector differential equation (the  auxiliary 2-surface $S$ is
arbitrary and static)
 $$
\frac{\partial}{\partial
\xi}\left(\mathbf{E}\times\mathbf{B}\right)=\mathfrak{F}, \ \ \ \xi\equiv ct,   \ \ \ \ \
\ \ (*)
$$
which is equivalent to
$$
\left(\mathrm{rot}\,\mathbf{E}+\frac{\partial\mathbf{B}}{\partial \xi}\right)\times
\mathbf{E}+ \mathbf{E}\mathrm{div}\,\mathbf{E}+
\left(\mathrm{rot}\,\mathbf{B}-\frac{\partial\mathbf{E}}{\partial \xi}\right)\times
\mathbf{B}+ \mathbf{B}\mathrm{div}\,\mathbf{B}=0.
$$
This last equation we write down in the following equivalent way:
$$
\left(\mathrm{rot}\,\mathbf{E}+\frac{\partial\mathbf{B}}{\partial \xi}\right)\times
\mathbf{E}+\mathbf{B}\mathrm{div}\,\mathbf{B}=
-\left[\left(\mathrm{rot}\,\mathbf{B}-\frac{\partial\mathbf{E}}{\partial
\xi}\right)\times\mathbf{B}+\mathbf{E}\mathrm{div}\,\mathbf{E}\right].\ \ \ \
(**)
 $$

The above relation (*) and its explicit forms we consider as mathematical
adequate of the so called electric-magnetic and magnetic-electric induction
phenomena in the charge free case since it represents these phenomena in
energy-momentum-change terms.

We recall that
it is usually assumed these induction phenomena to be described in classical
electrodynamics by the following well known integral equations
$$
\frac{d}{d\xi}\int_{S}*\tilde{g}(\mathbf{B})|_S=-
\int_{S}*\tilde{g}(\mathrm{rot}\mathbf{E})|_S
\ \ \ \text{(the Faraday induction law)},
$$
$$
\frac{d}{d\xi}\int_{S}*\tilde{g}(\mathbf{E})|_S=
\int_{S}*\tilde{g}(\mathrm{rot}\mathbf{B})|_S
\ \ \ \text{(the Maxwell displacement current law)},
$$
where $(...)|_S$ means restriction of the corresponding 2-form to the 2-surface
$S$.

We would like to note that these last Faraday-Maxwell relations have NO
{\it direct} energy-momentum change-propagation (i.e. force flow) nature, so
they could not be experimentally verified in a {\it direct} way. Our feeling is
that, in fact, they are stronger than needed. So, on the corresponding
solutions of these equations we'll be able to write down {\it formally
adequate} energy-momentum change expressions, but the correspondence of these
expressions with the experiment will crucially depend on the nature of these
solutions. As is well known, the nature of the free solutions (with no
boundary conditions) to Maxwell vacuum equations with spatially finite and
smooth enough initial conditions requires strong time-instability (the
corresponding  theorem for the D'Alembert wave equation which each component of
$\mathbf{E}$ and $\mathbf{B}$ must necessarily satisfy). And time-stability of
time-dependent vacuum solutions usually requires spatial infinity (plane
waves), which is physically senseless. Making calculations with spatially
finite parts of these spatially infinite solutions may be practically
acceptable, but from theoretical viewpoint assuming these equations for {\it
basic} ones seems not acceptable since the relation "time stable physical
object - exact free solution" is strongly violated.

Before to go further we write down the right hand side bracket expression of
$(**)$ in the following two equivalent ways:
$$
\left[\left(\mathrm{rot}\,\mathbf{B}+\frac{\partial\mathbf{(-E)}}{\partial
\xi}\right)\times \mathbf{B}+ \mathbf{(-E)}\mathrm{div}\,\mathbf{(-E)}\right];\,
$$
$$
\left[\left(\mathrm{rot}\,\mathbf{(-B)}+\frac{\partial\mathbf{E}}
{\partial\xi}\right)\times\mathbf{(-B)}+\mathbf{E}\mathrm{div}\,\mathbf{E}\right].
$$
These last two expressions can be considered as obtained from the left
hand side of the above relation $(**)$  under the substitutions
$(\mathbf{E},\mathbf{B})\rightarrow(\mathbf{B},\mathbf{-E})$ and
$(\mathbf{E},\mathbf{B})\rightarrow(\mathbf{-B},\mathbf{E})$ respectively.
Hence, the subsystem $(\mathbf{E},\mathbf{B})$ chooses as a partner
 $(\mathbf{-B},\mathbf{E})$, or $(\mathbf{B},\mathbf{-E})$.
We conclude that the subsystem
$(\mathbf{E},\mathbf{B},\alpha)$ will choose as partner-susbsystem
$(\mathbf{E},\mathbf{B},\alpha+\frac{\pi}{2})$ or
$(\mathbf{E},\mathbf{B},\alpha-\frac{\pi}{2})$.

We may summarize this nonrelativistic approach as follows:

{\bf A real free field consists of two interacting subsystems
$(\Sigma_1,\Sigma_2)$, and each
subsystem is described by two partner-fields
inside the $\theta(x,y,z;t)$-family
$$
(\mathcal{E},\mathcal{B})=
(\mathbf{E}\,\mathrm{cos}\,\theta- \mathbf{B}\,\mathrm{sin}\,\theta; \
\mathbf{E}\,\mathrm{sin}\,\theta+ \mathbf{B}\,\mathrm{cos}\,\theta),
$$
$\Sigma_1=(\mathcal{E}_{\theta_{1}},\mathcal{B}_{\theta_{1}}),
\Sigma_2=(\mathcal{E}_{\theta_{2}},\mathcal{B}_{\theta_{2}})$, giving the same
Maxwell stress-energy tensor, so the full stress energy tensor is the sum:
$M(\Sigma_1,\Sigma_2)=\frac12M(\Sigma_1)+\frac12M(\Sigma_2)$. Each
partner-field has interacting electric and magnetic constituents, and each
partner-field is determined by the other through $(\pm\frac{\pi}{2})$ -
rotation-like transformation.  Both partner-fields carry the same
stress-energy-momentum : $M(\Sigma_1)=M(\Sigma_2)$, and the field propagates in
space through minimizing the relation} $I_1^2+I_2^2\geqslant 0$. {\bf The
intrinsic dynamics of a free real time-dependent field establishes and
maintains local energy-momentum exchange partnership between the two
partner-fields, and since these partner-fields carry always the same
stress-energy, the allowed exchange is necessarily simultaneous and in equal
quantities, so, each partner-field conserves its energy-momentum during
propagation}.

\section{Internal interaction and evolution in energy-momentum terms}

In order to find how the local intercommunication between the constituents of
each subsystem, and the energy-momentum exchange between the two subsystems is
performed, we are going to interpret appropriately the above equation (**). Our
object of interest, representing the integrity of a real time dependent
electromagnetic field, is the couple
$\Big[(\mathbf{E},\mathbf{B});(\mathbf{-B},\mathbf{E})\Big]$ (the other case
$\Big[(\mathbf{E},\mathbf{B});(\mathbf{B},\mathbf{-E})\Big]$ is considered
analogically). In view of the above considerations our equations should
directly describe admissible energy-momentum exchange between these
recognizable two subsystems. Since these two subsystems are supposed to keep
their recognizabilty during spacetime propagation, we have to take in view
corresponding {\it admissible}, i.e. not leading to nonrecognizability but {\it
real}, changes. It is supposed, of course, that these admissible real changes
must be mathematicaly represented by tensor objects.

Let's denote such changes by
$D(\mathbf{E},\mathbf{B})$ and $D(\mathbf{-B},\mathbf{E})$ for each
partner-field, so, $D(\mathbf{E},\mathbf{B})$ will describe admissible changes
{\bf inside} the subsystem formally represented by $(\mathbf{E},\mathbf{B})$,
and $D(\mathbf{-B},\mathbf{E})$ will describe admissible changes {\bf inside}
the subsystem formally represented by $(\mathbf{-B},\mathbf{E})$. Their
{\bf admissible} nature now should formally mean that the values of the
corresponding self-couplings
$$ \mathfrak{P}
\left[D(\mathbf{E},\mathbf{B});(\mathbf{E},\mathbf{B})\right] \ \ \text{and} \ \
\ \mathfrak{P}\left[
D(\mathbf{-B},\mathbf{E});(\mathbf{-B},\mathbf{E})\right]
$$ might be zero, i.e. not essential. Any available cross-couplings
$$ \mathfrak{P}\left[
D(\mathbf{E},\mathbf{B});(\mathbf{-B},\mathbf{E})\right]
 \ \ \text{and}
\ \ \ \mathfrak{P}:
\left[D(\mathbf{-B},\mathbf{E});(\mathbf{E},\mathbf{B})\right]
 $$
will
formally represent {\bf admissible changes}, due to interaction between the two
subsystems. These last changes must be appropriately equalizable, guaranteing
in this way the required recognizability of each partner-field. From physical
point of view the values of all these couplings should have energy-momentum
change nature in order to be considered as really available.

The explicit forms of these self-couplings and cross-couplings are
suggested by equations (*), or (**).

Following this suggestion, the change object $D(\mathbf{E},\mathbf{B})$ for the
first partner-field $(\mathbf{E},\mathbf{B})$ we naturally define as
$$
D(\mathbf{E},\mathbf{B}):=
\left(\mathrm{rot}\mathbf{E}+ \frac{\partial\mathbf{B}}{\partial \xi}; \,
\mathrm{div}\mathbf{B}\right).
$$
The corresponding self-coupling
$\mathfrak{P}:D(\mathbf{E},\mathbf{B})\rightarrow(\mathbf{E},\mathbf{B})$
$$
\mathfrak{P}\left[D(\mathbf{E},\mathbf{B}); (\mathbf{E},\mathbf{B})\right]=
\mathfrak{P}\left[\left(\mathrm{rot}\mathbf{E}+ \frac{\partial\mathbf{B}}{\partial \xi};
\, \mathrm{div}\mathbf{B}\right); (\mathbf{E},\mathbf{B})\right]
$$
is suggested by the left hand side of the local energy-momentum
exchange relation (**), so we define it by :
$$
\mathfrak{P}\left[\left(\mathrm{rot}\mathbf{E}+
\frac{\partial\mathbf{B}}{\partial \xi}; \, \mathrm{div}\mathbf{B}\right);
(\mathbf{E},\mathbf{B})\right]:=
\left(\mathrm{rot}\,\mathbf{E}+\frac{\partial\mathbf{B}}{\partial \xi}\right)\times
\mathbf{E}+\mathbf{B}\mathrm{div}\,\mathbf{B}.
$$
For the second partner-field
$(-\mathbf{B},\mathbf{E})$, following the same procedure we obtain:
$$
\mathfrak{P}\left[D(\mathbf{-B},\mathbf{E}); (\mathbf{-B},\mathbf{E})\right]=
\mathfrak{P}\left[\left(\mathrm{rot}\mathbf{(-B)}+ \frac{\partial\mathbf{E}}{\partial
\xi}; \, \mathrm{div}\mathbf{E}\right); (\mathbf{-B},\mathbf{E})\right]
$$
$$
=\left(\mathrm{rot}\,\mathbf{(-B)}+\frac{\partial\mathbf{E}}{\partial \xi}\right)\times
\mathbf{(-B)}+\mathbf{E}\mathrm{div}\,\mathbf{E}=
\left(\mathrm{rot}\,\mathbf{B}-\frac{\partial\mathbf{E}}{\partial \xi}\right)\times
\mathbf{B}+\mathbf{E}\mathrm{div}\,\mathbf{E}.
$$
Hence, relation (**) takes the form
$$
\mathfrak{P}\left[D(\mathbf{E},\mathbf{B});
(\mathbf{E},\mathbf{B})\right]=- \mathfrak{P}\left[D(\mathbf{-B},\mathbf{E});
(\mathbf{-B},\mathbf{E})\right] .
$$

We turn now to the mutual energy-momentum exchange between the two
partner-fields $(\mathbf{E},\mathbf{B})\rightleftarrows
(\mathbf{-B},\mathbf{E})$, keeping in mind that each of the two partner
fields must keep its energy-momentum. The formal expressions are easy to
obtain. In fact, in the case
$[D(\mathbf{-B},\mathbf{E});(\mathbf{E},\mathbf{B})]$,
following the same coupling-procedure, we
have to couple the change object for the second partner-field given by
$$
D(\mathbf{-B},\mathbf{E}):= \left(\mathrm{rot}\mathbf{(-B)}+
\frac{\partial\mathbf{E}}{\partial \xi}; \, \mathrm{div}\mathbf{E}\right)
$$
with the first partner-field $(\mathbf{E},\mathbf{B})$. We obtain:
$$
\left(\mathrm{rot}\,(\mathbf{-B})+\frac{\partial\mathbf{E}}{\partial \xi}\right)\times
\mathbf{E}+\mathbf{B}\mathrm{div}\,\mathbf{E}=
-\left(\mathrm{rot}\,\mathbf{B}-\frac{\partial\mathbf{E}}{\partial \xi}\right)\times
\mathbf{E}+\mathbf{B}\mathrm{div}\,\mathbf{E} \ .
$$
In the reverse case we have to couple the change-object for the first
partner-field $(\mathbf{E},\mathbf{B})$ given by
$$
D(\mathbf{E},\mathbf{B}):= \left(\mathrm{rot}\mathbf{E}+
\frac{\partial\mathbf{B}}{\partial \xi}; \, \mathrm{div}\mathbf{B}\right)
$$
with the second partner-field $(\mathbf{-B},\mathbf{E})$. We obtain
$$
\left(\mathrm{rot}\,\mathbf{E}+\frac{\partial\mathbf{B}}{\partial
\xi}\right)\times (\mathbf{-B})+\mathbf{E}\mathrm{div}\,\mathbf{B}=
-\left(\mathrm{rot}\,\mathbf{E}+\frac{\partial\mathbf{B}}{\partial
\xi}\right)\times\mathbf{B}+\mathbf{E}\mathrm{div}\,\mathbf{B}.
$$

We recall now the {\it special kind of dynamical equilibrium} between the two
partner-fields, namely, {\it the two partner-fields necessarily carry always
the same stress-energy and momentum}:
$$
M^{ij}(\mathbf{E},\mathbf{B},\theta_1)=M^{ij}(\mathbf{E},\mathbf{B},\theta_2),
\ \ \ \mathbf{P}(\mathbf{E},\mathbf{B},\theta_1)
=\mathbf{P}(\mathbf{E},\mathbf{B},\theta_2),
 \text{where} \ \ \ |\theta_1-\theta_2|=\frac\pi 2,
$$
so, the mutual exchange is simultanious and in equal quantities.

The required permanent recognizability of each of the two
subsystems should be formally represented by the zero values of the two
self-couplings, and the local dynamical equilibrium between these two
subsystems should be formally represented by the the equal up
to a sign mutual couplings during any time period, so, the final equations
read:
\begin{center} \hfill\fbox{
    \begin{minipage}{0.97\textwidth}
\begin{center}
$$
\left(\mathrm{rot}\,\mathbf{E}+\frac{\partial\mathbf{B}}{\partial
\xi}\right)\times \mathbf{E}+\mathbf{B}\mathrm{div}\,\mathbf{B}=0,     
$$
$$
\left(\mathrm{rot}\,\mathbf{B}-\frac{\partial\mathbf{E}}{\partial      
\xi}\right)\times \mathbf{B}+\mathbf{E}\mathrm{div}\,\mathbf{E}=0,
$$
$$
\left(\mathrm{rot}\,\mathbf{E}+\frac{\partial\mathbf{B}}{\partial
\xi}\right)\times \mathbf{B}-\mathbf{E}\mathrm{div}\,\mathbf{B}+        
\left(\mathrm{rot}\,\mathbf{B}-\frac{\partial\mathbf{E}}{\partial \xi}\right)\times
\mathbf{E}-\mathbf{B}\mathrm{div}\,\mathbf{E}=0.
$$
\end{center}
\vskip 0.3cm
 \end{minipage}} \hfill \end{center}

The first equation describes admissible balance between the two constituents
$(\mathbf{E},\mathbf{B})$ inside the first subsystem, the second equation
describes admissible balance between the two constituents
$(\mathbf{-B},\mathbf{E})$ inside the second subsystem, and the third equation
establishes balance between the two subsystems, moreover, it  guarantees that
{\it the exchange of energy-momentum density between the two partner-fields is
{\bf simultaneous} and in {\bf equal} quantities}, i.e. a {\bf permanent
dynamical equilibrium} between the two partner-fields holds.

Note that, this double-field viewpoint on the field and the corresponding mutual
energy-momentum exchange described by these equations are essentially new
moments.

The above equations can be given the following form in terms of differential
forms. If $g$ is the euclidean metric let's introduce the following notations:
$$
\tilde{g}(\mathbf{E})=\eta, \ \ \tilde{g}(\mathbf{B})=\beta, \ \
\tilde{g}^{-1}(*\eta)=\bar{*\eta}, \ \ \tilde{g}^{-1}(*\beta)=\bar{*\beta}.
$$
Then we obtain
$$
\tilde{g}(\mathrm{rot}\mathbf{E}\times\mathbf{E})=i(\mathbf{E})\mathbf{d}\eta,\
\ \tilde{g}(\mathrm{rot}\mathbf{B}\times\mathbf{B})=i(\mathbf{B})\mathbf{d}\beta,
$$
$$
\tilde{g}(\mathrm{rot}\mathbf{E}\times\mathbf{B})=i(\mathbf{B})\mathbf{d}\eta, \ \
\tilde{g}(\mathrm{rot}\mathbf{B}\times\mathbf{E})=i(\mathbf{E})\mathbf{d}\beta
, $$ $$ \mathbf{E}\,\mathrm{div}(\mathbf{B})=i(\bar{*\eta})\mathbf{d}*\beta, \
\ \mathbf{B}\,\mathrm{div}(\mathbf{E})=i(\bar{*\beta})\mathbf{d}*\eta .
$$
Under these notations and relations the above three framed equations are
respectively equivalent to the following flows of vector fields across
differential forms expressions:
$$
i(\mathbf{E})\mathbf{d}\eta+i(\bar{*\beta})\mathbf{d}*\beta=
-*\left(\frac{\partial\beta}{\partial\xi}\wedge\eta\right)=-
i(\mathbf{E})\left(*\frac{\partial\beta}{\partial\xi}\right),
$$
$$
i(\mathbf{B})\mathbf{d}\beta+i(\bar{*\eta})\mathbf{d}*\eta=
*\left(\frac{\partial\eta}{\partial\xi}\wedge\beta\right)=
i(\mathbf{B})\left(*\frac{\partial\eta}{\partial\xi}\right),
$$
$$
i(\mathbf{B})\mathbf{d}\eta-i(\bar{*\eta})\mathbf{d}*\beta+
i(\mathbf{E})\mathbf{d}\beta-i(\bar{*\beta})\mathbf{d}*\eta=
i(\mathbf{E})\left(*\frac{\partial\eta}{\partial\xi}\right)-
i(\mathbf{B})\left(*\frac{\partial\beta}{\partial\xi}\right)
$$
Having in view the expressions for the Lie derivatives in Sec.2 we can rewrite
the expressions on the left of $"="$ in terms of Lie
derivatives as follows:
$$
L_{\mathbf{E}}\eta-\mathbf{d}\langle\eta,\mathbf{E}\rangle-
\big[L_{\bar{*\beta}}*\beta-\mathbf{d}\langle*\beta,\bar{*\beta}\rangle\big{]}=
-i(\mathbf{E})\left(*\frac{\partial\beta}{\partial\xi}\right)
$$
$$
L_{\mathbf{B}}\beta-\mathbf{d}\langle\beta,\mathbf{B}\rangle-
\big[L_{\bar{*\eta}}*\eta-\mathbf{d}\langle*\eta,\bar{*\eta}\rangle\big]=
i(\mathbf{B})\left(*\frac{\partial\eta}{\partial\xi}\right)
$$
$$
L_{\mathbf{E}}\beta-\mathbf{d}\langle\beta,\mathbf{E}\rangle+
L_{\mathbf{B}}\eta-\mathbf{d}\langle\eta,\mathbf{B}\rangle+
L_{\bar{*\eta}}*\beta-\mathbf{d}\langle*\beta,\bar{*\eta}\rangle+
L_{\bar{*\beta}}*\eta-\mathbf{d}\langle*\eta,\bar{*\beta}\rangle
$$
$$
=i(\mathbf{E})\left(*\frac{\partial\eta}{\partial\xi}\right)
-i(\mathbf{B})\left(*\frac{\partial\beta}{\partial\xi}\right).
$$
These last relations show that the couplings may be expressed, in fact, as local
self-flows and mutual-flows of the mathematical images of the physical electric
and magnetic vector field constituents of the field across the differential
form images of their admissible nonzero changes.

\section{Some properties of the equations and their solutions}
Consider the second equation and replace
$(\mathbf{E},\mathbf{B})$ acording to
$$
(\mathbf{E},\mathbf{B})
\rightarrow(a\mathbf{E}-b\mathbf{B},b\mathbf{E}+a\mathbf{E}),
$$
where $(a,b)$ are two constants.
After the corresponding computation we obtain
$$
a^2\left[
\left(\mathrm{rot}\,\mathbf{B}- \frac{\partial \mathbf{E}}{\partial
\xi}\right) \times\mathbf{B}+\mathbf{E}\,\mathrm{div}\,\mathbf{E}\right]+
b^2\left[\left(\mathrm{rot}\,\mathbf{E}+
\frac{\partial \mathbf{B}}{\partial \xi}\right)
\times\mathbf{E}+\mathbf{B}\,\mathrm{div}\,\mathbf{B}\right]+
$$
$$
+ab\left[
\left(\mathrm{rot}\,\mathbf{E}+
\frac{\partial \mathbf{B}}{\partial \xi}\right)
\times\mathbf{B}-\mathbf{E}\,\mathrm{div}\,\mathbf{B}+
\left(\mathrm{rot}\,\mathbf{B}-
\frac{\partial \mathbf{E}}{\partial \xi}\right)
\times\mathbf{E}-\mathbf{B}\,\mathrm{div}\,\mathbf{E}\right]=0.
$$
Since the constants $(a,b)$ are arbitrary the other two equations
follow. The same property holds with respect to any of the three equations.

	{\bf Corollary.} The system of the three equations is invariant with
respect to the transformation
$$
(\mathbf{E},\mathbf{B})
\rightarrow(a\mathbf{E}-b\mathbf{B},b\mathbf{E}+a\mathbf{E}).
$$
Writing down this transformation in the form
\begin{eqnarray*}
(\mathbf{E},\mathbf{B})\rightarrow &(\mathbf{E}',\mathbf{B}')          
=(\mathbf{E},\mathbf{B}).\alpha(a,b)
=(\mathbf{E},\mathbf{B})\begin{Vmatrix} a & b \\ -b & a \end{Vmatrix}\\&
=(a\mathbf{E}-b\mathbf{B}, b\mathbf{E}+a\mathbf{B}), \
a=const, \ b=const,
\end{eqnarray*}
we get a "right action" of the matrix $\sigma(a,b)$ on the solutions.
 The new solution $(\mathbf{E}',\mathbf{B}')$ has energy and momentum densities
equal to the old ones multiplied by $(a^2+b^2)$. Hence, the space of all
solutions factors over the action of the group of matrices of the kind
\[
\sigma(a,b)= \begin{Vmatrix} a & b \\ -b & a \end{Vmatrix},
\quad (a^2+b^2)\neq 0
\]
in the sense, that the corresponding classes are determined by the value of
$(a^2+b^2)$.

All such matrices with nonzero determinant form a group
with respect to the usual matrix product. A special property of this group is
that it represents the symmetries of the canonical complex structure in
$\mathbb{R}^2$.

Clearly, all solutions to Maxwell pure field equations are solutions to
our nonlinear equations, we shall call these solutions linear, and
will not further be interested of them, we shall concentrate our attention on
those solutions of our equations which satisfy the conditions
$$
\mathrm{rot}\,\mathbf{E}+\frac{\partial\mathbf{B}}{\partial \xi}\neq 0,\quad
\mathrm{rot}\,\mathbf{B}-\frac{\partial\mathbf{E}}{\partial \xi}\neq 0,\quad
\mathrm{div}\,\mathbf{E}\neq 0,\quad \mathrm{div}\,\mathbf{B}\neq 0.
$$
These solutions we call further nonlinear.

We consider now some properties of the {\it nonlinear} solutions.
\vskip 0.3cm
$\bf 1.$ Among the nonlinear solutions  there
are no constant ones.
\vskip 0.3cm
$\bf 2.$\ $\mathbf{E}.\mathbf{B}=0;$ This is obvious, no proof is needed.
\vskip 0.3cm
$\bf 3.$ The following relations are also obvious:
$$\left(\mathrm{rot}\,\mathbf{E}+ \frac{\partial\mathbf{B}}{\partial
\xi}\right).\mathbf{B}=0; \ \ \left(\mathrm{rot}\,\mathbf{B}-
\frac{\partial\mathbf{E}}{\partial \xi}\right).\mathbf{E}=0.
$$
\noindent
\vskip 0.3cm
{\bf 4.} It is elementary to see from the last two relations that the
classical Poynting energy-momentum balance equation follows.
 \vskip 0.3cm
{\bf 5.} $\mathbf{E}^2=\mathbf{B}^2$.

In order to prove this let's take the scalar product of the first
equation from the left by ${\mathbf B}$.  We obtain

$$
{\mathbf B}.\Biggl\{\left(\mathrm{rot}{\mathbf E}+
\frac{\partial {\mathbf B}}{\partial \xi}\right)
\times {\mathbf E}\Biggr\}+{\mathbf B}^2\mathrm{div}{\mathbf B}=0. \ \ \ \ \ (1)
$$
Now, multiplying the third equation from the left by ${\mathbf E}$ and having
in view $\mathbf{E}.\mathbf{B}=0$, we obtain
$$
{\mathbf E}.\Biggl\{\left(\mathrm{rot}{\mathbf E}+
\frac{\partial {\mathbf E}}{\partial \xi}\right) \times {\mathbf
B}\Biggr\}-{\mathbf E}^2\mathrm{div}{\mathbf B}=0.
$$
 This last relation is equivalent
to
$$
-{\mathbf B}.\Biggl\{\left(\mathrm{rot}{\mathbf E}+ \frac{\partial {\mathbf
B}}{\partial \xi}\right) \times {\mathbf E} \Biggr\}- {\mathbf
E}^2\mathrm{div}{\mathbf B}=0.\ \ \ \\ (2)
$$
Now, summing up $(1)$ and $(2)$, in view of $\mathrm{div}{\mathbf B}\neq 0$,
we come to the desired relation.

Properties {\bf 2.} and {\bf 5.} say that all nonlinear solutions are {\it
null fields}, i.e. the two well known relativistic invariants
$I_1=\mathbf{B}^2-\mathbf{E}^2$ and $I_2=2\mathbf{E}.\mathbf{B}$ of the field
are zero, and this property leads to optimisation of the inequality
$I_1^2+I_2^2\geqslant 0$ (recall the eigen properties of Maxwell stress tensor),
 which, in turn, guarantees $\theta(x,y,z;t)$-invariance of $I_1=I_2=0$.

\vskip 0.3cm
{\bf 6.} The {\it helicity} property:
$$
\mathbf{E}.\left(\mathrm{rot}\,\mathbf{E}+
\frac{\partial\mathbf{B}}{\partial \xi}\right)- \mathbf{B}.
\left(\mathrm{rot}\,\mathbf{B}- \frac{\partial\mathbf{E}}{\partial \xi}\right)=
\mathbf{E}.\mathrm{rot}\mathbf{E}-\mathbf{B}.\mathrm{rot}\mathbf{B}=0.
$$

\noindent To prove this property we first multiply (vector product) the third
equation from the right by $\mathbf{E}$, make use of the relation
$(\mathbf{X}\times\mathbf{Y})\times\mathbf{Z}=\mathbf{Y}(\mathbf{X}.\mathbf{Z})-
\mathbf{X}(\mathbf{Y}.\mathbf{Z})$ and recall property {\bf 2}., then
multiply (scalar product) from the left by $\mathbf{B}$, recall again
property {\bf 2}, and, finally, recall property {\bf 5.}

Property {\bf 6.} suggests the following consideration. If $\mathbf{V}$ is an
arbitrary vector field on $\mathbb{R}^3$ then the quantity
$\mathbf{V}.\mathrm{rot}\mathbf{V}$ is known as {\it local helicity} and its
integral over the whole (compact) region occupied by $\mathbf{V}$ is known as
{\it integral helicity}, or just as {\it helicity} of $\mathbf{V}$. Hence,
property {\bf 6.} says that the electric and magnetic constituents of a
nonlinear solution generate the same helicities. If we consider (through the
euclidean metric $g$) the 1-form $\tilde{g}(\mathbf{E})$ and denote by
$\mathbf{d}$ the exterior derivative on $\mathbb{R}^3$, then
$$
\tilde{g}(\mathbf{E})\wedge\mathbf{d}\tilde{g}(\mathbf{E})=
\mathbf{E}.\mathrm{rot}\mathbf{E}\,dx\wedge dy\wedge dz,
$$
so, the zero helicity says that the 1-form $\tilde{g}(\mathbf{E})$ defines a
completely integrable Pfaff system:
$\tilde{g}(\mathbf{E})\wedge\mathbf{d}\tilde{g}(\mathbf{E})=0$. The nonzero
helicity says that each of the 1-forms $\tilde{g}(\mathbf{E})$ and
$\tilde{g}(\mathbf{B})$ defines non-integrable 1-dimensional  Pfaff system, so
the nonzero helicity defines corresponding curvature. Therefore the equality
between the $\mathbf{E}$-helicity and the $\mathbf{B}$-helicity suggests to
consider the corresponding integral helicity
$$
\int_{\mathbb{R}^3}\tilde{g}(\mathbf{E})\wedge\mathbf{d}\tilde{g}(\mathbf{E}) =
\int_{\mathbb{R}^3}\tilde{g}(\mathbf{B})\wedge\mathbf{d}\tilde{g}(\mathbf{B})
$$
(when it takes finite nonzero values) as a measure of the spin properties
of the solution.

We specially note that the equality of the local helicities
defined by $\mathbf{E}$ and $\mathbf{B}$ holds also, as it is easily seen from
the above relation, for the solutions of the linear Maxwell vacuum equations,
but appropriate solutions giving well defined and time independent integral
helicities in this case are missing. The next property shows that our nonlinear
solutions admit such appropriate solutions giving finite constant integral
helicities.

 \vskip 0.3cm {\bf 7.}\ \ Example of nonlinear solution(s):
\begin{align*} &\mathbf{E}=\left[\phi(x,y,\xi\pm z)
\mathrm{cos}(-\kappa\frac{z}{\mathcal{L}_o}+const), \, \phi(x,y,\xi\pm
z)\mathrm{sin}(-\kappa\frac{z}{\mathcal{L}_o}+const),\,0\right];\\ &\mathbf{B}=\left[\pm
\phi(x,y,\xi\pm z)\, \mathrm{sin}(-\kappa\frac{z}{\mathcal{L}_o}+const),\,\mp
\phi(x,y,\xi\pm z) \mathrm{cos}(-\kappa\frac{z}{\mathcal{L}_o}+const),\,0\right],
\end{align*}
where $\phi(x,y,\xi\pm z)$ is an arbitrary positive function,
$0<\mathcal{L}_o<\infty$ is an arbitrary positive constant
with physical dimension of length, and $\kappa$ takes values $\pm1$.
Hence, we are allowed to choose the function $\phi$ to have compact
3d-support, and since the energy density of this solution is $\phi^2dx\wedge
dy\wedge dz$ and the total energy is conserved according to property {\bf 4.},
these solutions will describe time-stable and space propagating with the
speed of light finite field objects carrying finite integral energy.

Modifying now the corresponding helicity 3-forms to
$$
\frac{2\pi
\mathcal{L}_o^2}{c}\tilde{g}(\mathbf{E})\wedge\mathbf{d}\tilde{g}(\mathbf{E})=
\frac{2\pi
\mathcal{L}^2_o}{c}\tilde{g}(\mathbf{B})\wedge\mathbf{d}\tilde{g}(\mathbf{B}),
$$
then the corresponding 3d integral gives $\kappa TE$, where $\kappa=\pm 1$,
$T=2\pi\mathcal{L}_o/c$ and $E=\int{\phi^2}dx\wedge dy\wedge dz$ is the integral
energy of the solution.

\section{Scale factor} We consider the vector
fields $$ \vec{\mathcal{F}}=\mathrm{rot}\,\mathbf{E}+ \frac{\partial
\mathbf{B}}{\partial \xi}+ \frac{\mathbf{E}\times\mathbf{B}}
{|\mathbf{E}\times\mathbf{B}|}\,\mathrm{div}\,\mathbf{B},   
$$
$$
\vec{\mathcal{M}}=\mathrm{rot}\,\mathbf{B}-
\frac{\partial {\mathbf E}}{\partial \xi}-                   
\frac{\mathbf{E}\times\mathbf{B}}
{|\mathbf{E}\times\mathbf{B}|}\,\mathrm{div}\,\mathbf{E},
$$
defined by a nonlinear solution.

It is obvious that on the solutions of Maxwell's vacuum equations
$\vec{\mathcal{F}}$ and $\vec{\mathcal{M}}$ are equal to zero. Note also that
under the transformation
$(\mathbf{E},\mathbf{B})\rightarrow (\mathbf{-B},\mathbf{E})$
we get
$\vec{\mathcal{F}}\rightarrow
-\vec{\mathcal{M}}$ and $\vec{\mathcal{M}}\rightarrow \vec{\mathcal{F}}$.

We shall consider now the relation between  $\vec{\cal F}$ and $\mathbf{E}$,
and between $\vec{\cal M}$ and $\mathbf{B}$ on the nonlinear solutions of our
equations assuming that $\vec{\cal F}\neq 0$ and $\vec{\cal M}\neq 0$.

Recalling $\mathbf{E}.\mathbf{B}=0$ we obtain
$$
({\mathbf E}\times {\mathbf
B})\times {\mathbf E}= -{\mathbf E}\times ({\mathbf E}\times {\mathbf B})=
-[{\mathbf E}({\mathbf E}.{\mathbf B})-
{\mathbf B}({\mathbf E}.{\mathbf E})]={\mathbf B}({\mathbf E}^2),
$$
and since $|{\mathbf E}\times {\mathbf B}|
=|\mathbf{E}||\mathbf{B}||\mathrm{sin}(\mathbf{E},\mathbf{B})|=
 {\mathbf E}^2={\mathbf B}^2$, we get
$$
\vec{\cal F}\times {\mathbf E}=
\left(\mathrm{rot}{\mathbf E}+\frac{\partial {\mathbf B}}{\partial \xi}\right)
\times {\mathbf E}+{\mathbf B}\mathrm{div}{\mathbf B}=0,
$$
according to our first nonlinear equation.

In the same way, in accordance with our second nonlinear equation,
we get $\vec{\cal M}\times {\mathbf B}=0$. In other words, on
the nonlinear solutions we obtain that $\vec{\cal F}$ is co-linear to
${\mathbf E}$ and $\vec{\cal M}$ is co-linear to ${\mathbf B}$.  Hence, we
can write the relations
$$
\vec{\cal F}=f_1.{\mathbf E},\ \ \vec{\cal M}=f_2.{\mathbf B},        
$$
where $f_1$ and $f_2$ are two functions, and of course, the
interesting cases are $f_1\neq 0,\infty;\ f_2\neq 0,\infty$.
Note that the physical dimension of $f_1$ and $f_2$ is the
reciprocal to the dimension of coordinates, i.e.
$[f_1]=[f_2]= [length]^{-1}$.

Note also that $\vec{\cal F}$ and $\vec{\cal M}$ are mutually orthogonal:
$\vec{\cal F}.\vec{\cal M}=0$.

We shall prove now that $f_1=f_2$. In fact, making use of the same formula
for the double vector product, used above, we easily obtain
$$
\vec{\cal F}\times {\mathbf B}+\vec{\cal M}\times {\mathbf E}=
$$
$$
=\left(\mathrm{rot}{\mathbf E}+\frac{\partial {\mathbf B}}{\partial
\xi}\right)
\times {\mathbf B}+
\left(\mathrm{rot}{\mathbf B}-\frac{\partial {\mathbf E}}{\partial \xi}\right)
\times {\mathbf E}-{\mathbf E}\mathrm{div}{\mathbf B}
-{\mathbf B}\mathrm{div}{\mathbf E}=0,
$$
in accordance with our third nonlinear equation.
Therefore,
$$
\vec{\cal F}\times {\mathbf B}+\vec{\cal M}\times {\mathbf E}=
$$
$$
=f_1{\mathbf E}\times {\mathbf B}+f_2{\mathbf B}\times {\mathbf E}=
(f_1-f_2){\mathbf E}\times {\mathbf B}=0.
$$
The assertion follows.

The relation $|\vec{\cal F}|=|\vec{\cal M}|$ is now obvious.

Note that the two relations $|\vec{\cal F}|^2=|\vec{\cal M}|^2$ and
$\vec{\cal F}.\vec{\cal M}=0$  and the duality correspondence $(\vec{\cal
F},\vec{\cal M})\rightarrow (-\vec{\cal M},\vec{\cal F})$
 make us consider $\vec{\cal F}$ and $\vec{\cal M}$
as nonlinear analogs of $\mathbf{E}$ and $\mathbf{B}$ respectively.

These considerations suggest to introduce the quantity
$$
\mathcal{L}(\mathbf{E},\mathbf{B})=
\frac{1}{|f_1|}=\frac{1}{|f_2|}=
\frac{|\mathbf{E}|}{|\vec{\mathcal{F}}|}=                      
\frac{|\mathbf{B}|}{|\vec{\mathcal{M}}|},
$$
which we call {\it scale factor}.
Note that the physical dimension of  $\mathcal{L}$ is {\it
length}. Hence, {\it every} nonlinear solution defines its own {\it scale
factor} and, concequently, the nonlinear solutions factorize with respect to
$\mathcal{L}$. It seems natural to connect the constant $\mathcal{L}_o$ in the above
given family of solutions with the so introduced scale factor.
Assuming $\mathcal{L}=\mathcal{L}_o=const$, this could be done in the following
way.

A careful look at the solutions above shows that at a given moment, e.g. $t=0$,
the finite spatial support of the function $\phi$ is built of continuous sheaf
of nonintersecting helices along the coordinate $z$. Every such helix has a
special length parameter $b=\lambda/2\pi$ giving the straight-line advance
along the external straight-line axis (the coordinate $z$ in our case) for a
unit angle, and $\lambda$ is the $z$-distance between two equivalent points on
the same helix. So, we may put $\lambda=2\pi\mathcal{L}_o=const$, hence, the
$z$-size of the solution may, naturally, be bounded by $2\pi\mathcal{L}_o$, so,
$\mathcal{L}_o$ should coincide with the radius of the projection of the helix
on the plane $(x,y)$.

Consider now a nonlinear solution with integral energy $E$ and scale factor
$\mathcal{L}_o=const$. Since this solution shall propagate translationally in
space with the speed of light $c$, we may introduce corresponding time period
$T=2\pi\mathcal{L}_o/c$, and define the quantity $\mathfrak{h}=E.T$, having
physical dimension of "action". The temptation to separate a class of
solutions, requiring $\mathfrak{h}$ to be equal to the Planck constant $h$ is
great, especially if this $T$ can be associated with some helix-like
real periodicity during propagation!

\section{Conclusion}
The viewpoint we paid due respect in this paper consists in the following.
Physical reality demonstrates itself through creating spatially finite
entities called by us {\it physical objects}. These entities show two aspect
nature: {\it physical appearance} and {\it time existence and recognizability}.
The physical appearance of a physical object is understood as corresponding
{\it stress-strain abilities}, presenting the spatial structure of the
object. The {\it time existence and recognizability} require {\it survival
abilities}, which demonstrate the {\it dynamical appearance} of the oblect
through building stress-energy-momentum characteristics, formally represented
by the tensor $T^{ij}$ and momentum vector $\mathbf{P}$, on one hand, and
corresponding {\it acting instruments}, i.e., local flows, formally represented
by the tensor constituents of the divergence $\nabla_{i}T^{ij}$ and time
drivative of $\mathbf{P}$, on the other hand.

In trying to formalize this understanding, the basic idea we followed in this
paper was to pass from linear to nonlinear equations in charge free classical
electrodynamics through giving explicit physical sense of the equations as
local stress-energy-momentum relations. This view on the subject naturally
oriented our attention to the quantities carrying the necessary information -
the corresponding stress-energy tensors, their divergences, and the
corresponding self flows and mutual flows . The existing knowledge about the
structure and internal dynamics of free electromagnetic field objects made us
assume the notion for {\it two partner-fields internal structure}. Each of
these two partner-fields consists of two constituents, each partner-field is
able to carry local momentum and to allow local "intercomunication" between its
two constituents in presence of its partner-field. The two subsystems carry
equal local energy-momentum densities, and realize local mutual energy exchange
without available interaction energy. Moreover, they strictly respect each
other: the exchange is simultaneous and in equal quantities, so, both
partner-fields keep their identity and recognizability. The corresponding
internal dynamical structure appropriately unifies translation and rotation
through unique space-time propagation with the fundamental velocity. All
Maxwell solutions are duly respected. The new solutions are time-stable, they
admit finite spatial support, and minimize the relation $I_1^2+I_2^2 \geq 0$.
These spatially finite solutions are of photon-like nature: they are
time-stable, the propagate as a whole with the velocity of light in vacuum, they
demonstrate intrinsically compatible translational-rotational dynamical
structure, they carry finite energy-momentum and intrinsically determined
integral characteristic $\mathfrak{h}$ of action nature through appropriate
scale factor $\mathcal{L}_o=const$. Their integral energy $E$ satisfies
relation of the form identical to the Planck formula $E.T=\mathfrak{h}$.

\vskip 1cm
{\bf References}
\vskip 0.4cm

[1]. {\bf M. Born, L. Infeld}, {\it Nature}, {\bf 132}, 970 (1932)

[2]. {\bf M. Born, L.Infeld}, {\it Proc.Roy.Soc.}, {\bf A 144}, 425 (1934)

[3]. {\bf W. Heisenberg, H. Euler}, {\it Zeit.Phys.}, {\bf 98}, 714 (1936)

[4]. {\bf M. Born}, {\it Ann. Inst. Henri Poincare}, {\bf 7}, 155-265 (1937).

[5]. {\bf J. Schwinger}, {\it Phys.Rev}. ,{\bf 82}, 664 (1951).

[6]. {\bf H. Schiff}, {\it Proc.Roy.Soc.} {\bf A 269}, 277 (1962).

[7]. {\bf J. Plebanski}, {\it Lectures on Nonlinear Electrodynamics}, NORDITA,
Copenhagen, 1970.

[8]. {\bf G. Boillat}, {\it Nonlinear Electrodynamics: Lagrangians and
Equations of Motion}, \newline J.Math.Phys. {\bf 11}, 941 (1970).

[9]. {\bf B. Lehnert, S. Roy}, {\it Extended Electromagnetic Theory}, World
Scientific, 1998.

[10]. {\bf D.A. Delphenich}, {\it Nonlinear Electrodynamics and QED},
arXiv:hep-th/0309108, (good review article).

[11]. {\bf B. Lehnert}, {\it A Revised Electromagnetic Theory with Fundamental
Applications}, Swedish Physic Arhive, 2008.

[12]. {\bf D. Funaro}, {\it Electromagnetsm and the Structure of Matter},
Worldscientific, 2008; also: {\it From photons to atoms}, arXiv: gen-ph/1206.3110
(2012).

[13]. {\bf E. Schrodinger}, {\it Contribution to Born's new theory of
electromagnetic feld}, Proc. Roy. Soc. Lond. {\bf A 150}, 465 (1935).

[14]. {\bf G. Gibbons, D. Rasheed}, {\it Electric-magnetic duality rotations in
non-linear electrodynamics}, Nucl. Phys. {\bf B 454} 185 (1995) hep-th/9506035.

[15] {\bf R. Kerner, A.L. Barbosa, D.V. Gal'tsov}, {\it Topics in Born-Infeld
Electrodynamics}, arXiv: hep-th/0108026 v2

[16]. {\bf J. Marsden, A. Tromba}, {\it Vector Calculus}, fifth edition, W.H.
Freeman and Company, 2003.

[17]. {\bf M. Forger, C. Paufler, H. Reomer}, {\it The Poisson Bracket for
Poisson Forms in Multisymplectic Field Theory}, arXiv: math-ph/0202043v1

 \end{document}